\documentclass[prx,twocolumn,superscriptaddress,nofootinbib,showpacs,notitlepage]{revtex4-2}
\usepackage{amsmath}
\usepackage{amsfonts}
\usepackage{amssymb}
\usepackage{bbm}
\usepackage[utf8]{inputenc}
\usepackage{braket}
\usepackage{environ}
\usepackage{graphicx,color,dsfont}
\usepackage{booktabs}
\usepackage{floatrow}
\usepackage[caption=false,label font={bf,normalsize}]{subfig}
\usepackage[dvipsnames]{xcolor}
\usepackage[all]{xy}
\usepackage{relsize}
\usepackage{hyperref}
\usepackage{accents}
\hypersetup{linktocpage}
\definecolor{darkblue}{RGB}{0,0,127} 
\definecolor{darkgreen}{RGB}{0,150,0}
\hypersetup{colorlinks=true,citecolor=darkblue,linkcolor=darkblue, 
	filecolor=red, urlcolor=darkblue,pdftitle={Three-dimensional quantum cellular automata from chiral semion surface topological order and beyond},
pdfauthor={Wilbur Shirley, Yu-An Chen, Arpit Dua, Tyler D. Ellison, Nathanan Tantivasadakarn, and Dominic J. Williamson}}
\usepackage{lmodern}

\newcommand\thickbar[1]{\accentset{\rule{.5em}{.5pt}}{#1}}

\newtheorem{theorem}{Theorem}[section]

\newtheorem{conjecture}{Conjecture}[section]

\newtheorem{corollary}{Corollary}[section]

\usepackage{lipsum}

\newtheorem{definition}{Definition}[section]

\makeatletter
\def\@opargbegintheorem#1#2#3{\trivlist
   \item[]{\bfseries #1\ #2\ (#3)} \itshape}
\makeatother

\NewEnviron{eqs}{%
\begin{equation}\begin{split}
    \BODY
\end{split}\end{equation}
}

\newcommand{\ZZ}{\mathbb{Z}}
\begin{document}

\title{Three-dimensional quantum cellular automata from chiral semion surface topological order and beyond}

\author{Wilbur Shirley}
\email{wshirley@ias.edu}
\affiliation{School of Natural Sciences, Institute for Advanced Study, Princeton, NJ 08540, USA}
\affiliation{Department of Physics and Institute for Quantum Information and Matter, California Institute of Technology, Pasadena, CA 91125, USA}

\author{Yu-An Chen}
\affiliation{Department of Physics, Condensed Matter Theory Center, and Joint Quantum Institute, University of Maryland, College Park, MD 20742, USA}
 
\author{Arpit Dua}
\affiliation{Department of Physics and Institute for Quantum Information and Matter, California Institute of Technology, Pasadena, CA 91125, USA}
 
\author{Tyler~D. Ellison}
\affiliation{Department of Physics, University of Washington, Seattle, WA 98195, USA}
\affiliation{Department of Physics, Yale University, New Haven, CT 06511, USA}

\author{Nathanan Tantivasadakarn}
\affiliation{Department of Physics, Harvard University, Cambridge, MA 02138, USA}

\author{Dominic~J. Williamson}
\thanks{Current Address: Centre for Engineered Quantum Systems, School of Physics,
University of Sydney, Sydney, New South Wales 2006, Australia}
\affiliation{Stanford Institute for Theoretical Physics, Stanford University, Stanford, CA 94305, USA}

\begin{abstract}
We construct a novel three-dimensional quantum cellular automaton (QCA) based on a system with short-range entangled bulk and chiral semion boundary topological order. We argue that either the QCA is nontrivial, \textit{i.e.}, not a finite-depth circuit of local quantum gates, or there exists a two-dimensional commuting projector Hamiltonian realizing the chiral semion topological order (characterized by $U(1)_2$ Chern-Simons theory). Our QCA is obtained by first constructing the Walker-Wang Hamiltonian of a certain premodular tensor category of order four, then condensing the deconfined bulk boson at the level of lattice operators. We show that the resulting Hamiltonian hosts chiral semion surface topological order in the presence of a boundary and can be realized as a non-Pauli stabilizer code on qubits, from which the QCA is defined. The construction is then generalized to a class of QCAs defined by non-Pauli stabilizer codes on ${2^n}$-dimensional qudits that feature surface anyons described by $U(1)_{2^n}$ Chern-Simons theory. Our results support the conjecture that the group of nontrivial three-dimensional QCAs is isomorphic to the Witt group of non-degenerate braided fusion categories.
\end{abstract}

\maketitle

\tableofcontents

\section{Introduction}

Quantum cellular automata (QCAs) are locality-preserving unitary operators on quantum many-body lattice systems.\footnote{For infinite-sized systems, it is more precise to regard QCAs as locality-preserving operator algebra automorphisms. On a finite-sized system, a locality-preserving unitary defines such an automorphism by conjugation.} They originally arose in the context of quantum simulation and as a model for quantum computation~\cite{QC1,QC2,QC3,QC4,QC5}. However, in recent years, QCAs have seen wide-ranging applications from discretized quantum field theories~\cite{QFT1,QFT2} to the classification of Floquet phases~\cite{PoChiral,PoRadical,PotterMorimoto17,PotterVishwanathFidkowski18,Zhang2021classification,Glorioso21} and tensor network unitary operators~\cite{IgnacioCirac2017,Sahinoglu2018,GongSunderhaufSchuchCirac20,Piroli21Fermionic,Piroli2020}, entanglement growth in quantum dynamics~\cite{Nahum,GongPiroliCirac21,Ranard20,GongNahumPiroli21}, and the construction of symmetry-protected topological states and their anomalous boundaries ~\cite{JonesMetlitski21,HFH2020}. Beyond such applications, QCAs represent a fundamental class of mathematical objects in the quantum many-body setting, meshing the notions of unitarity and locality. Thus, they merit a thorough investigation in their own right. 

In any spatial dimension, there are two very natural classes of QCAs. The first is that of finite-depth quantum circuits (FDQCs) constructed from local gates, which are unitary by definition and locality-preserving by merit of their finite depth. The second is that of discrete translations. In fact, in one spatial dimension the index theory of Gross, Nesme, Vogts, and Werner fully classifies the set of QCAs in terms of translations and FDQCs \cite{GNVW12}. Moreover this classification has been extended to the two-dimensional case via techniques of dimensional reduction \cite{FH20,Haah21}. A natural question to ask is whether there exist QCAs in higher dimensions that lie beyond these two ``trivial" classes, \textit{i.e.}, ones that are not equivalent to the composition of a translation with an FDQC.

Remarkably, Haah, Fidkowski, and Hastings \cite{HFH18} identified a QCA in three spatial dimensions, which they conjectured to be ``nontrivial" in the sense above. The QCA is best understood in terms of an intriguing connection with the class of ($3+1$)D exactly-soluble lattice models known as Walker-Wang models \cite{WW12}. In particular, the QCA is derived from the Walker-Wang model based on the 3-fermion modular tensor category (MTC).\footnote{For our intents and purposes, a modular tensor category is an anyon theory with the property that for every anyon $a$ there exists an anyon $a'$ that braids nontrivially with $a$.} The associated Walker-Wang model is short-range entangled in the bulk and harbors 3-fermion surface topological order in the presence of an exposed boundary \cite{BCFV14}. The QCA defined in Ref.~\cite{HFH18} has the key property that it disentangles the 3-fermion Walker-Wang model -- mapping it to a sum of single site Pauli $Z$ operators. A subsequent work of Haah \cite{Haah21} introduced a class of QCAs that disentangle the eigenstates of exactly-soluble ($3+1$)D lattice Hamiltonians whose surface topological states are characterized by nonzero chiral central charge and anyonic excitations with fusion group $\mathbb{Z}_p$ for odd prime $p$.

The common thread among these three-dimensional QCAs, which are conjectured to be nontrivial, is the property that they disentangle commuting projector Hamiltonians that host chiral surface topological orders in the presence of a boundary. As argued in Ref.~\cite{HFH18}, if one assumes that such a QCA is trivial, \textit{i.e.}, if it is equivalent to an FDQC times a translation, then it follows that one can construct a commuting projector Hamiltonian realizing the chiral surface topological order in a strictly two-dimensional system. More specifically, a bulk FDQC can always be truncated to define an FDQC on a system with an exposed boundary. For instance, the truncated FDQC can be defined by simply throwing away all gates of the bulk circuit that are not fully supported within a given region. Thus, if the disentangling QCA was trivial, one could truncate it and apply it to the commuting projector Hamiltonian with surface topological order. This would trivialize the bulk, hence disentangling it from the boundary and leaving behind a strictly 2D commuting projector boundary Hamiltonian for the surface topological order. This contradicts the widely held belief that chiral topological orders cannot be realized by commuting projector Hamiltonians \cite{kitaev2006anyons,FK_CommProj2019,SK_CommProj2020}, thus providing strong evidence for the nontriviality of these novel QCAs.

In this paper, we introduce a new three-dimensional QCA, referred to as $\alpha_1$, with the property that it disentangles the eigenstates of a novel lattice Hamiltonian $H_1$, which hosts the chiral semion anyon theory in the presence of a boundary. We conjecture, via the argument outlined above, that $\alpha_1$ is a nontrivial QCA. We emphasize that, although $H_1$ shares similar properties to the Walker-Wang model based on the chiral semion theory \cite{vKBS13}, \textit{i.e.}, vanishing correlation length, exact solubility, short-range entanglement, 
% unique ground state, 
and boundary terminations with chiral semion surface topological order, $H_1$ is not the chiral semion Walker-Wang model. Our identification of $\alpha_1$ relies on the notion of a \textit{locally flippable separator} -- a class of exactly-solvable models introduced in Ref.~\cite{HFH18}. According to Theorem~11.4 of Ref.~\cite{HFH18}, locally flippable separators are in one-to-one correspondence with QCAs. The key feature of $H_1$, in contrast with the chiral semion Walker-Wang model, is that it satisfies the criteria required of a locally flippable separator (reviewed in Sec.~\ref{sec:QCA}), thus it corresponds to a QCA.

We recall that the chiral semion topological order is characterized by the Abelian MTC $\{1,s\}$, whose single nontrivial quasiparticle $s$ is a semion (topological spin $\theta=i$) with $\mathbb{Z}_2$ fusion rules. It is equivalently realized by $U(1)_2$ Chern-Simons theory which arises as the low energy description of the $\nu=1/2$ bosonic Laughlin fractional quantum Hall state. The Hamiltonian $H_1$ can be readily generalized to a series of exactly-solvable, short-range entangled Hamiltonians $H_n$, whose boundary terminations host surface topological orders respectively characterized by $U(1)_{2^n}$ Chern-Simons theory. Likewise, the QCA $\alpha_1$ generalizes to a class of QCAs $\alpha_n$ which respectively disentangle the Hamiltonian $H_n$ to a sum of single site Pauli operators.

The identification of this class of QCAs is notable for two reasons. First, it differs from previous constructions of nontrivial three-dimensional QCAs in that it is not a manifestly Clifford QCA, \textit{i.e.}, one that maps generalized Pauli operators to local products of Pauli operators. Second, and more importantly, our findings mark a significant step toward a classification of nontrivial three-dimensional QCAs. As we have discussed, there is an evident link between nontrivial QCAs in 3D and chiral surface topological orders in 2D. The identification of 3D QCAs corresponding to chiral semion 2D order, and more generally that of $U(1)_{2^n}$, fill a gap in this emerging story. 

In fact, the novel QCAs we introduce, combined with those for qudits of odd prime dimension described in Ref.~\cite{Haah21}, likely exhaust all classes of nontrivial QCAs that disentangle exactly-solvable 3D lattice models with Abelian surface topological order. To be precise, the set of QCAs modulo FDQCs and translations in any particular spatial dimension is known to form an Abelian group. In three dimensions, it has been conjectured that this group contains a subgroup isomorphic to a mathematical object known as the \textit{Witt group of metric groups} \cite{Davydov2013a,Davydov2013b}. According to this conjecture, this group is generated by the QCA of Ref.~\cite{Haah21} and those of the present work. In particular this would imply that each $\alpha_n$ has order 8, the first such examples of QCAs with finite order greater than 4, and moreover that the family of QCAs $\alpha_n$ generate a $\mathbb{Z}_8\oplus\mathbb{Z}_2$ group. In fact, $\alpha_1$ and $\alpha_2$ are sufficient to generate this group; $\alpha_1$ corresponds to the generator of $\mathbb{Z}_8$ and $\alpha_1\alpha_2^{-1}$ the generator of $\mathbb{Z}_2$. Note that the $U(1)_4$ surface topological order corresponding to $\alpha_2$ is equivalent to the $\nu=2$ state in the Kitaev's sixteen-fold way \cite{kitaev2006anyons}. We remark that the QCA $\alpha_1\alpha_2^{-1}$ corresponds to surface topological order characterized by $K=\left(\begin{smallmatrix}2&0\\0&-4\end{smallmatrix}\right)$ Chern-Simons theory, which is nonchiral but also does not admit a gapped boundary \cite{Levin2013Protected,LL14}. Thus the argument for nontriviality of this QCA relies on the assumption that this topological order, or more generally any non-gappable Abelian topological order, cannot be realized by a commuting projector Hamiltonian. Indeed, this has been conjectured to be the case \cite{LL14}.

The paper is organized as follows. In Sec.~\ref{sec:QCA}, we discuss the main technical result of the paper: the construction of $H_1$ and the disentangling QCA $\alpha_1$, as well as the boundary Hamiltonian with chiral semion surface topological order. In Sec.~\ref{sec:N} we generalize the construction to arbitrary $H_n$ and $\alpha_n$. Sec.~\ref{sec:Witt} reviews a series of conjectures relating the total group of nontrivial three-dimensional QCAs to the so-called categorical Witt group\footnote{Also known as the Witt group of non-degenerate braided fusion categories.}~\cite{Davydov2013a,Davydov2013b}, which takes into account the possibility of disentangling QCAs for lattice models with non-Abelian surface topological order. We conclude with a discussion in Sec.~\ref{sec:discussion}.

\section{Construction of a quantum cellular automaton from chiral semion surface topological order}
\label{sec:QCA}

In this section, we construct the Hamiltonian $H_1$ on a system of qubits and a QCA that maps $H_1$ to a trivial sum of single-qubit Pauli $Z$ operators. Furthermore, we construct a Hamiltonian with a boundary that hosts chiral semion surface topological order, and which is a sum of mutually commuting terms, hence exactly-solvable. $H_1$ is defined on a cubic lattice with one qubit per edge $e$. In terms of operators $\hat{B}_p$ associated to each plaquette $p$, $H_1$ is:
\begin{equation}
    H_1=-\sum_p\hat{B}_p.
\end{equation}
The operators $\hat{B}_p$, which are defined in Sec.~\ref{sec:transformation}, belong to the Clifford group. Moreover, they each square to the identity. As mentioned in the introduction, $H_1$ is similar to the chiral semion Walker-Wang model in the sense that it has no ground state degeneracy under periodic boundary conditions, but admits a surface termination on lattices with boundary that harbor chiral semion topological order. The key property of $H_1$ that differentiates it from the bona fide Walker-Wang model is that the collection of $\hat{B}_p$ terms constitutes a \textit{locally flippable separator} on a tensor product Hilbert space, in the sense of Ref.~\cite{HFH18}. 

\begin{definition}[\cite{HFH18}]
A {locally flippable $\mathbb{Z}_2$ separator} is an indexed set of operators $B_a$ (separators) and $F_a$ (flippers), each supported in a finite-radius disk, satisfying:
\begin{enumerate}
    \item{$B_a^2=1$}
    \item{$[B_a,B_b]=\{F_a,B_a\}=[F_a,B_b]=0$ for $a\neq b$.}
    \item{For any assignment $a\mapsto\omega(a)=\pm1$, the space of states $\ket{\psi}$ such that $B_a\ket{\psi}=\omega(a)\ket{\psi}$ for all $a$ is one-dimensional.}
\end{enumerate}
\label{def:LFS}
\end{definition}

\begin{figure*}[htbp]
    \centering
    \includegraphics[width=\textwidth]{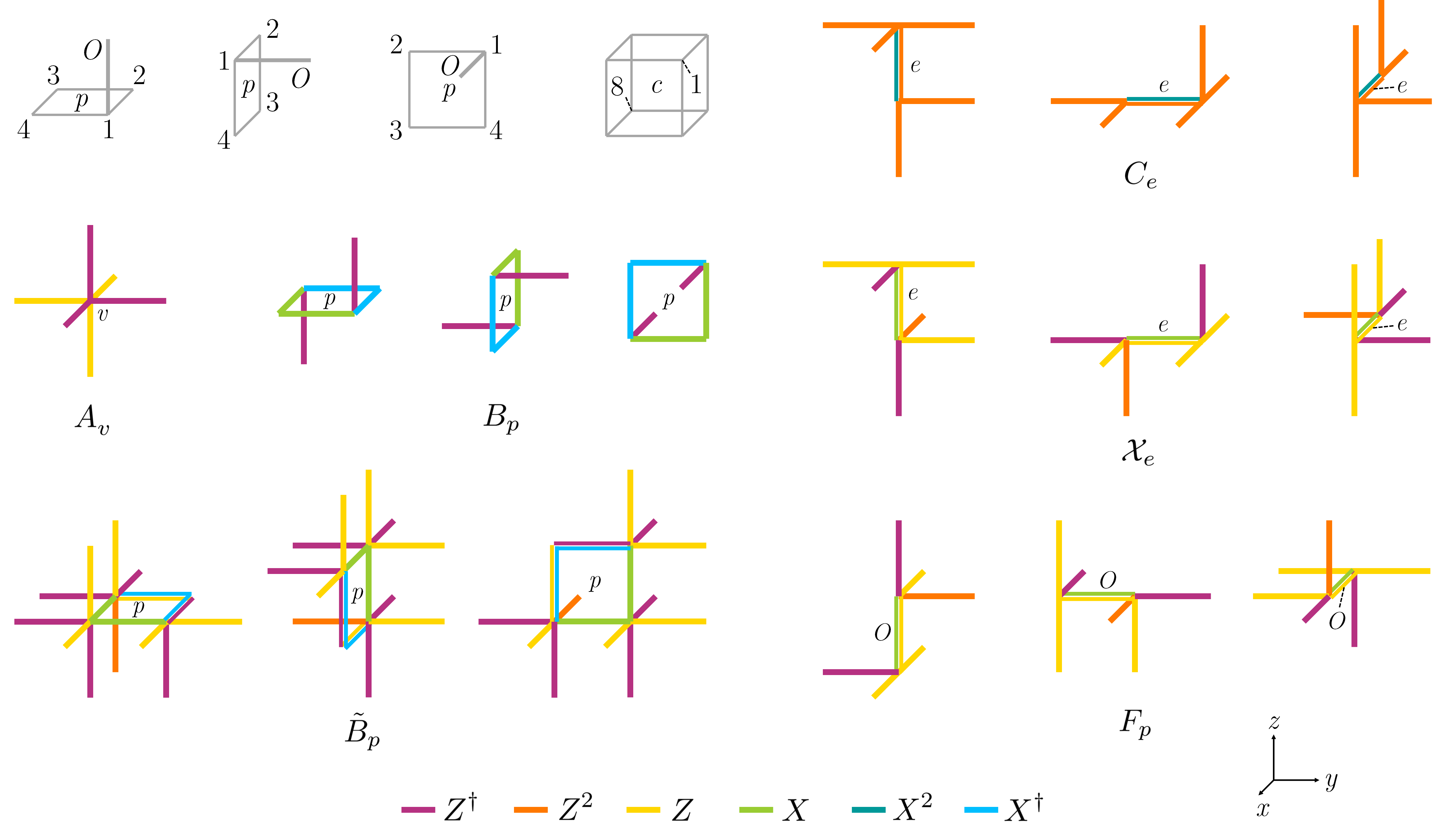}
    \caption{Vertex term $A_v$, plaquette terms $B_p,$ modified plaquette terms $\tilde{B}_p$, boson hopping operators $C_e$, semion hopping operators $\mathcal{X}_e$, and flippers $F_p$, defined with respect to the $O$ edge of $p$. Each operator is a tensor product of $\mathbb{Z}_4$ Pauli operators over the indicated qudits, with a legend shown at the bottom. Edges with two colors represent a product of the two operators, with $Z$ type operators preceding $X$ type. All operators have an overall phase of $+1$ except the $\mathcal{X}_e$ operators which have an overall $e^{-i\pi /4}$ phase. A polynomial representation of these operators is given in Appendix \ref{sec:poly}.}
    \label{fig:H1}
\end{figure*}

\noindent In Sec.~\ref{sec:transformation}, we define flippers $\hat{F}_p$ for the terms $\hat{B}_p$ of the $H_1$ such that $\hat{B}_p$ and $\hat{F}_p$ constitute a locally flippable $\mathbb{Z}_2$ separator. Theorem II.4 of Ref.~\cite{HFH18} then guarantees the existence of a QCA that disentangles the eigenstates of $H_1$ on a system with periodic boundaries. In particular, Ref.~\cite{HFH18} showed that, given a set of locally supported flippers $\hat{F}_p$, there always exists an alternative set of locally supported flippers $\hat{F}'_p$ satisfying:
\begin{align}
    \left(\hat{F}'_p\right)^2=1,\qquad\left[\hat{F}'_p,\hat{F}'_{p'}\right]=0.
\end{align}
The QCA $\alpha_1$ is then defined by mapping separators and modified flippers to Pauli $Z$ and Pauli $X$ operators, respectively:
\begin{equation}
    \alpha_1\left(\hat{B}_p\right) = X_O,\qquad \alpha_1\left(\hat{F}'_p\right) = Z_O,
\end{equation}
where the $O$ edge is defined relative to plaquette $p$ as in Fig.~\ref{fig:H1}. Ref.~\cite{HFH18} provides an algorithm for constructing the modified flippers $\hat{F}'_p$, which we do not explicitly carry out in this work due to the complexity of the original flippers $\hat{F}_p$.

\begin{figure}[t]
    \centering
    \includegraphics[scale=.55,trim={6.3cm 8.5cm 6cm 9cm},clip]{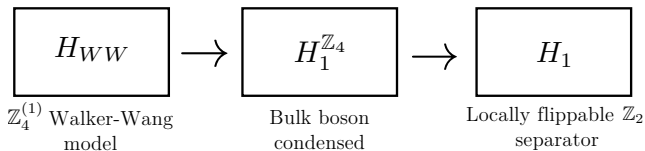}
    \caption{The construction of the locally flippable $\mathbb{Z}_2$ separator $H_1$ begins with a Walker-Wang model $H_{WW}$ based on the $\mathbb{Z}_4^{(1)}$ premodular anyon theory (Sec.~\ref{sec:WW}). $H_{WW}$ has an order-two bulk boson, which, when condensed, produces the Pauli stabilizer model $H_1^{\mathbb{Z}_4}$ defined on four-dimensional qudits (Sec.~\ref{sec:cond}). $H_1$ is then obtained by mapping each four-dimensional qudit to a pair of qubits, applying an FDQC composed of $\text{CNOT}$ gates, and projecting into a particular subspace (Sec.~\ref{sec:transformation}). $H_1$ is a locally flippable separator and defines the QCA $\alpha_1$.}
    \label{fig: H1flowchart}
\end{figure}

The construction of $H_1$ proceeds as follows (see Fig.~\ref{fig: H1flowchart}). In Sec.~\ref{sec:WW}, we begin with the Walker-Wang model $H_{WW}$ based on the $\mathbb{Z}_4^{(1)}$ premodular tensor category described in Ref.~\cite{ParsaThesis}.\footnote{Here, by a premodular tensor category, we mean an anyon theory in which at least one of the anyons braids trivially with every anyon in the theory.} $\mathbb{Z}_4^{(1)}$ is an Abelian theory with fusion group $\mathbb{Z}_4$, trivial $F$-symbols, and $R$-symbols $R^{a,b}_{a+b}=\exp{i\pi\frac{ab}{2}}$~\cite{ParsaThesis}. The generating anyon is therefore a semion, and the order two anyon is a boson with trivial braiding statistics. The $\mathbb{Z}_4^{(1)}$ Walker-Wang model thus has bulk 3D topological order equivalent to that of a 3D $\mathbb{Z}_2$ gauge theory, and $\mathbb{Z}_4$ surface semions, two of which fuse into the bulk deconfined boson. Moreover, $H_1$ is naturally expressed as a $\mathbb{Z}_4$ Pauli stabilizer code. In Sec.~\ref{sec:cond} we subsequently condense the deconfined $\mathbb{Z}_2$ boson in this model, yielding a $\mathbb{Z}_4$ Pauli stabilizer code Hamiltonian dubbed $H_1^{\mathbb{Z}_4}$, which is short-range entangled in the bulk but hosts chiral semion topological order in the presence of a boundary. Finally, in Sec.~\ref{sec:transformation} a unitary transformation followed by a projection yields the exactly-solvable non-Pauli Hamiltonian $H_1$ with these same properties. In Sec.~\ref{sec:boundary}, we construct the explicit boundary Hamiltonians for $H_{WW}$, $H_1^{\mathbb{Z}_4}$, and $H_1$.

\subsection{$\mathbb{Z}_4^{(1)}$ Walker-Wang model}
\label{sec:WW}

We begin with the $\mathbb{Z}_4^{(1)}$ Walker-Wang model. The system is composed of four-dimensional qudits on each edge of a cubic lattice, characterized by generalized Pauli operators $Z$ and $X$ obeying the $\mathbb{Z}_4$ clock and shift algebra
\begin{equation}
    Z^4=X^4=1\qquad ZX=iXZ.
\end{equation}
The Hamiltonian takes the form
\begin{equation}
    H_{WW}=-\sum_vA_v-\sum_p{B}_p+\text{h.c.}
\end{equation}
It contains a term $A_v$ associated to each cubic lattice vertex $v$, and a term $B_p$ associated to each plaquette $p$. These operators are defined in Fig.~\ref{fig:H1}. These Hamiltonian terms are mutually commuting, unfrustrated products of generalized $\mathbb{Z}_4$ Pauli operators, hence $H_{WW}$ is exactly-solvable, and moreover, a $\mathbb{Z}_4$ Pauli stabilizer code~\cite{NielsenChuang}. As mentioned, $H_{WW}$ exhibits bulk $\mathbb{Z}_2$ topological order, equivalent to that of the 3D toric code or 3D $\mathbb{Z}_2$ gauge theory. It is instructive to explicitly verify some of the defining properties of this topological order. (It may also be helpful to consider a coupled layer construction of $H_{WW}$ as described in Appendix~\ref{app: coupledlayer}).

First we compute the ground state degeneracy of $H_{WW}$ when it is placed on a spatial three-torus, in other words on a lattice with periodic boundary conditions. Let us first fix a notational convention. In defining $B_p$ we have implicitly endowed each plaquette $p$ with the positive normal orientation. Flipping the orientation of a plaquette is equivalent to Hermitian conjugation of the corresponding plaquette operator: $B_{\bar{p}}=B^\dagger_p$ where $\bar{p}$ ($p$) is negatively (positively) oriented. In the following, we consider products $\prod_{p\in c}$ over the plaquettes $p$ of an elementary lattice cube $c$. For such products, it is understood that each $p$ has the outward-facing orientation.\footnote{Explicitly, $\prod_{p\in c}B_p=B_1B_2B_3(B_4B_5B_6)^\dagger$ where $1,2,3$ are the faces adjacent to vertex 1, and $4,5,6$ are the faces adjacent to vertex 8.} With this in mind, we observe that the $B_p$ plaquette operators satisfy the relation
\begin{equation}
    \prod_{p\in c}B_p=A_1A_8,
\end{equation}
where the $1$ and $8$ vertices of cube $c$ are defined as in Fig. \ref{fig:H1}. To facilitate the ground state degeneracy calculation, and in anticipation of the boson condensation procedure in the following section, we define modified plaquette operators
\begin{equation}
    \tilde{B}_p\equiv B_p(A_1A_3A_4)^\dagger
\end{equation}
where the $1$, $3$, and $4$ vertices are indicated relative to $p$ in Fig.~\ref{fig:H1}. These operators satisfy the alternative property
\begin{equation}
    \prod_{p\in c}\tilde{B}_p=A_1^2
    \label{eq:Bprel}
\end{equation}
and hence
\begin{equation}
    \prod_{p\in c}\tilde{B}_p^2=1,
    \label{eq:Bp2rel}
\end{equation}
which greatly simplifies the counting of stabilizer generators and relations. We also define the modified stabilizer Hamiltonian
\begin{equation}
    \tilde{H}_{WW}=-\sum_vA_v-\sum_p\tilde{B}_p+\text{h.c.}
\end{equation}
Clearly, $H_{WW}$ and $\tilde{H}_{WW}$ generate the same stabilizer group, and thus have coinciding ground spaces. We work with $\tilde{H}_{WW}$ as it is more convenient.
Suppose there are $N$ sites in the lattice. Then there are $3N$ qudits and $4N$ stabilizer generators. There are $N-1$ independent relations owing to Eq.~\eqref{eq:Bprel}, and a single relation between all $A_v$ terms, namely $\prod_vA_v=1$. Each of these $N$ relations is of order 4. On a three-torus, there are also three independent relations of order 2, of the form of Eq.~\eqref{eq:Bp2rel} except that $c$ now belongs to one of the three classes of topologically nontrivial lattice 2-cycles. Therefore, the ground state degeneracy of $H_{WW}$ under periodic boundary conditions is $2^3=8$, as expected for 3D $\mathbb{Z}_2$ topological order.

Next, we identify the string operators that create pairs of the deconfined $\mathbb{Z}_2$ boson excitations that characterize 3D $\mathbb{Z}_2$ topological order. As we show below, for the Hamiltonian $\tilde{H}_{WW}$ these bosons correspond to isolated excitations of a single $A_v$ vertex term. Since local operators can only excite pairs of $A_v$ terms, such isolated excitations represent a deconfined, fractionalized quasiparticle. Naively, we may expect a product of $X^2$ operators along a string of edges to create a pair of bosons at its endpoints, since such an operator commutes with all $A_v$ terms except the pair at either end of the string, with which it anticommutes. However, such an operator also excites $\tilde{B}_p$ plaquette operators along the length of the string, hence it is not the correct string operator for the deconfined boson. Instead, we observe that, owing to Eq.~\eqref{eq:Bp2rel}, a product of $\tilde{B}_p^2$ operators over an open surface is a loop operator supported near the boundary of the surface. In fact, such a loop operator corresponds precisely to the motion of a deconfined $\mathbb{Z}_2$ boson around the loop. To see this, we define a set of short string operators $C_e$ for each edge $e$ of the lattice (Fig. \ref{fig:H1}), which satisfy the relations
\begin{equation}
    \tilde{B}^2_p=\prod_{e\in p}C_e,\qquad C_e^2=1.
    \label{eq:Cerel}
\end{equation}
Moreover, these operators satisfy the commutation relations
\begin{equation}
    C_eA_v=\begin{cases}-A_vC_e&\text{if $v\in e$}\\+A_vC_e&\text{if $v\not\in e$}\end{cases},
    \quad \left[C_e,\tilde{B}_p\right]=[C_e,C_{e'}]=0,
\end{equation}
implying that $C_e$ creates a pair of deconfined bosons at the two vertices of edge $e$. Therefore, topological string operators can be constructed by taking products of $C_e$ operators over all the edges of a given path. Such an operator commutes with all terms of $\tilde{H}_{WW}$ except the $A_v$ vertex terms at the endpoints of the path. Closed string operators of this form that wind around topologically nontrivial cycles of the spatial three-torus correspond to logical operators acting nontrivially on the eight-fold degenerate ground space under periodic boundary conditions.\footnote{This operator is also a topological string operator for the unmodified Hamiltonian $H_{WW}$, but it excites $B_p$ terms in the vicinity of each endpoint in addition to a single $A_v$ excitation.} In the following section, these $C_e$ operators play a key role in condensing the deconfined boson.

Finally, we identify the fractionalized loop-like excitation of 3D $\mathbb{Z}_2$ topological order. The vertex term $A_v$ can be interpreted as the elementary closed membrane operator for this loop excitation, and larger membrane operators enclosing a region $R$ are given by the product $\prod_{v\in R}A_v$. The anticommutation between $A_v$ and open string operators with an endpoint at $v$ is consistent with a $\pi$ braiding statistic between this loop-like excitation and the deconfined $\mathbb{Z}_2$ boson quasiparticle. Although $A_v$ has order four as a stabilizer generator, the loop-like excitation fuses with itself into a set of trivial (non-fractionalized) excitations in the vicinity of the loop, thus it obeys $\mathbb{Z}_2$ fusion rules as expected.\footnote{This can be verified in light of Eq.~\eqref{eq:BpXe}.}

We remark that when $H_{WW}$ is placed on a lattice with boundary, there is a natural boundary Hamiltonian that can be defined which maintains the $\mathbb{Z}_4$ Pauli stabilizer code nature of the model, and hosts a chiral semionic surface anyon. In this model a pair of surface semions fuses into the deconfined bulk boson. A detailed discussion of the boundary physics is reserved for Sec.~\ref{sec:boundary}.

\subsection{Condensing the bulk $\mathbb{Z}_2$ boson}
\label{sec:cond}

Next, we define a new $\mathbb{Z}_4$ Pauli stabilizer Hamiltonian $H_1^{\mathbb{Z}_4}$ that physically represents a system obtained by condensing the deconfined $\mathbb{Z}_2$ boson excitation of $H_{WW}$. The Hilbert space remains that of one four-dimensional qudit per edge of a cubic lattice, and the Hamiltonian takes the form
\begin{equation}
    H_1^{\mathbb{Z}_4}=-\sum_p\tilde{B}_p+\text{h.c.}-\sum_eC_e.
\end{equation}
This Hamiltonian has the property that its stabilizer group contains all $C_e$ operators along with all products of terms of $H_{WW}$ that commute with $C_e$. The inclusion of the boson creation operators $C_e$ in the stabilizer group implies condensation of bosons in the ground state, whereas the exclusion of the $A_v$ vertex terms corresponds to the resulting confinement of the fractionalized loop excitations of $H_{WW}$.\footnote{Note that $A_v^2$ commutes with $C_e$ for all $v,e$, and it is included in the stabilizer group since it is generated by $\tilde{B}_p$ terms.} To verify that this Hamiltonian lies in the condensed phase, \textit{i.e.}, with no bulk topological order, let us compute the ground state degeneracy of the model under periodic boundary conditions. On a lattice with $N$ sites, there are $3N$ qudits, $6N$ stabilizer generators, and $6N$ relations of order 2 of the form Eq.~\eqref{eq:Cerel}. There are no other relations, hence there is no ground state degeneracy, as expected.

As mentioned, $H_{WW}$ can be defined on a lattice with boundary such that it hosts a semionic surface anyon. Condensing the bulk boson destroys the bulk topological order but leaves intact a chiral semion surface topological order. As discussed in Sec.~\ref{sec:boundary}, there is a natural Pauli stabilizer code boundary termination of $H_1^{\mathbb{Z}_4}$ exhibiting this surface order.

The essential property of $H_1^{\mathbb{Z}_4}$ is that the $\tilde{B}_p$ terms constitute a locally flippable $\mathbb{Z}_2$ separator of the constrained Hilbert space in which all constraints of the form $C_e=1$ are satisfied. To demonstrate this, we identify flipper operators $F_p$, defined in Fig. \ref{fig:H1}, which satisfy
\begin{equation}
    \left\{F_p,\tilde{B}_p\right\}=\left[F_p,\tilde{B}_{p'}\right]=\big[F_p,C_e\big]=0,
    \label{eq:Fprop}
\end{equation}
where $p'\neq p$. Property 1 of Definition \ref{def:LFS} is satisfied due to Eq.~\eqref{eq:Cerel}, whereas property 3 is satisfied by a simple counting argument: The $C_e=1$ Hilbert space has dimension $2^E$, where $E$ is the number of edges, and there are $2^E$ common eigenstates of all $\tilde{B}_p$ operators. These eigenstates are uniquely indexed by their $\pm1$ eigenvalues and can be obtained by successively acting on the ground state with the flippers. We note that the form of the flippers $F_p$ is motivated by the discussion in Sec.~\ref{sec:boundary}.

To define a QCA via the theorem of Ref.~\cite{HFH18}, however, it is necessary to identify a locally flippable separator of a tensor product Hilbert space. In the next section, we explicitly transform the $C_e=1$ constrained Hilbert space into a tensor product Hilbert space by unitarily mapping the set of $C_e$ operators to single-qubit Pauli operators on unique qubit degrees of freedom. The separator Hamiltonian $H_1$ is obtained by projecting these qubits out from the transformed $H_1^{\mathbb{Z}_4}$.

\subsection{Obtaining a locally flippable $\mathbb{Z}_2$ separator}
\label{sec:transformation}

\begin{figure*}[htbp]
    \centering
    \includegraphics[width=\textwidth]{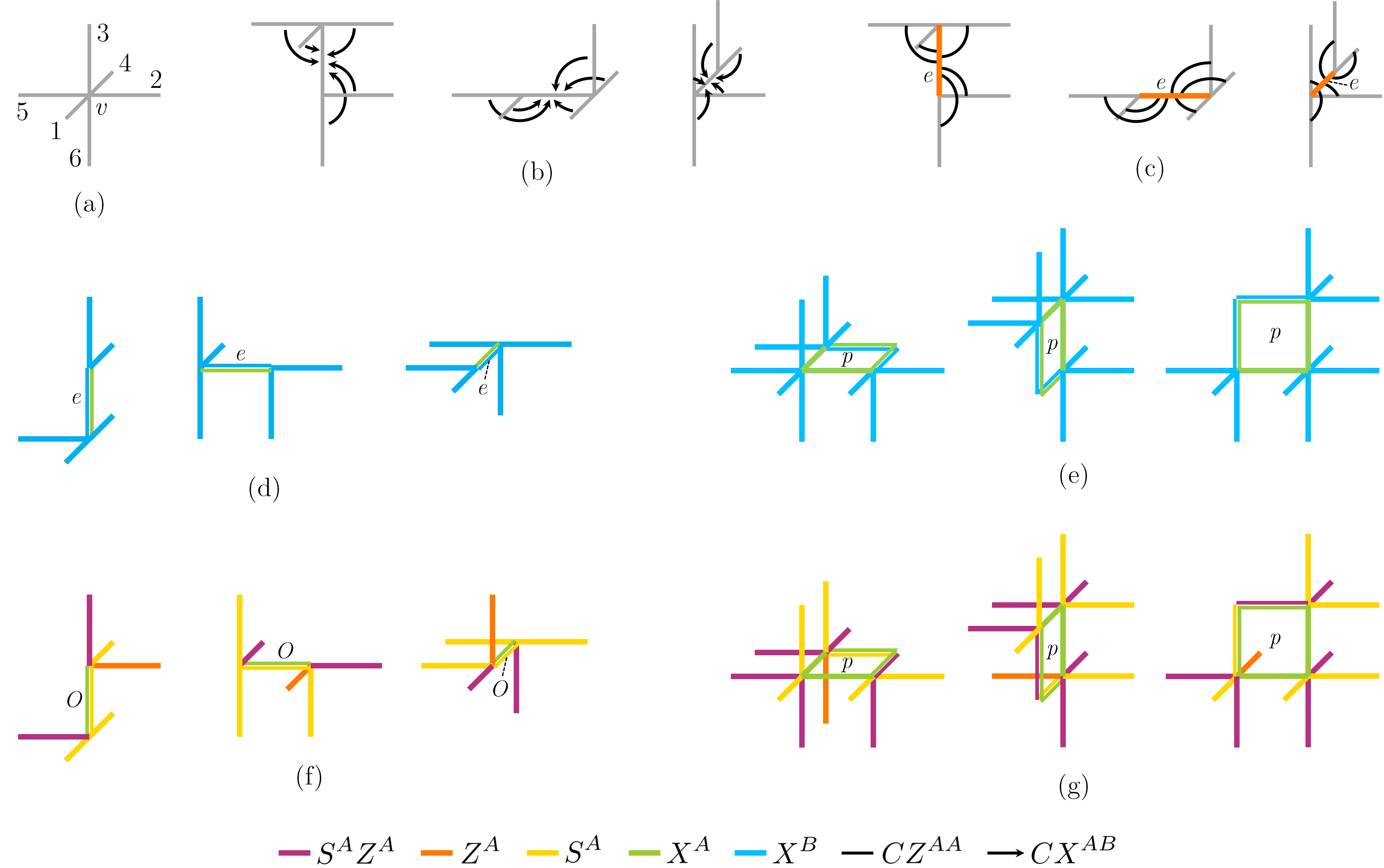}
    \caption{(a) Definition of the edges $e_1$ through $e_6$ with respect to vertex $v$. (b) A depiction of the operator $\prod_vU_v$. For a given edge, the figure shows all CNOT gates whose target qubit lies on that edge. Each gate is represented by an arrow; the control is qubit $A$ at the tail and the target is qubit $B$ at the head. (c) The operator $G_e$. (d) The images of $X_e^A$ and (e) $\prod_{e\in p}X_e^A$ under $U$. (f) The operator $\hat{F}_p^0$, defined with respect to the $O$ edge of $p$. (g) The operator $\hat{B}_p^0$. Each operator is a product of single-qubit and two-qubit gates over the indicated qubits, with a legend shown at the bottom. A black arc between two edges represents $CZ^{AA}$. Edges with two colors represent a product of the two operators, with operators diagonal in the Pauli $Z$ basis preceding Pauli $X$ operators.}
    \label{fig:circuit}
\end{figure*}

We now carry out this procedure to obtain the locally flippable separator Hamiltonian $H_1$, which is short-range entangled in the bulk but harbors chiral semion surface topological order in the presence of a boundary. The first step is to express $H_1^{\mathbb{Z}_4}$ in terms of qubit degrees of freedom rather than four-dimensional qudits. The four-dimensional qudit on each edge is instead regarded as a pair of qubits via the operator algebra automorphism
\begin{equation}
\begin{split}
    \thickbar{X}\to X^ACZ^{AB}\qquad
    &\thickbar{Z}\to X^BS^A\\
    \thickbar{X}^2\to Z^B\qquad
    &\thickbar{Z}^2\to Z^A
    \label{eq:auto}
\end{split}
\end{equation}
where $\thickbar{Z}$ and $\thickbar{X}$ represent $\mathbb{Z}_4$ Pauli operators and $A$ and $B$ label the two qubits~\cite{TwistedFFP}. Here we define
\begin{equation}
    S^A\equiv i^{\frac{1-Z^A}{2}}\qquad
    CZ^{AB}\equiv (-1)^{\frac{1-Z^A}{2}\frac{1-Z^B}{2}}.
\end{equation}
Under this mapping, the $C_e$, $\tilde{B}_p$, and $F_p$ operators acquire a new form as operators on the qubit Hilbert space -- however, by a slight abuse of notation, we continue to use the same symbols for the new operators.

The next step is to perform the circuit $U$ of controlled-$X$ gates defined as follows:
\begin{equation}
    U=\prod_e{CX}^{AB}_{e}\prod_vU_v,
\end{equation}
where the products are over all edges and vertices respectively, and with $U_v$ given by
\begin{equation}
    \begin{split}
        U_v=\hspace{.1cm}&CX^{AB}_{23}CX^{AB}_{63}CX^{AB}_{16}CX^{AB}_{26}CX^{AB}_{56}\\
        &CX^{AB}_{12}CX^{AB}_{52}CX^{AB}_{15}CX^{AB}_{35}CX^{AB}_{45}\\
        &CX^{AB}_{31}CX^{AB}_{41}CX^{AB}_{24}CX^{AB}_{34}CX^{AB}_{64}
    \end{split}
\end{equation}
where ${CX}_{ij}^{AB}$ denotes a controlled-$X$ gate with the $A$ qubit on edge $i$ as the control and the $B$ qubit on edge $j$ as the target. The numbering of edges with respect to vertex $v$ is as depicted in Fig. \ref{fig:circuit}(a), and the operator $\prod_vU_v$ is shown in Fig. \ref{fig:circuit}(b). Note that all of the $CX$ gates comprising $U$ mutually commute since the targets are all $B$ qubits, hence $U$ has finite depth. 
Recall that ${CX}_{ij}$ acts as
\begin{equation}
\begin{split}
    X_i\to X_iX_j\qquad X_j\to X_j \qquad &Z_i\to Z_i\qquad Z_j\to Z_iZ_j\\CZ_{ij}\to Z_iCZ_{ij}\qquad &CZ_{jk}\to CZ_{ik}CZ_{jk}.
\end{split}
\end{equation}
By construction $U$ maps each $C_e$ operator to a single Pauli $Z^B_e$, effectively disentangling each $B$ qubit from the rest of the system. $U$ acts trivially on $X^B_e$, $S^A_e$, and $Z^A_e$, and nontrivially on $X^A_e$ and $CZ^{AB}_e$. The image of $X^A_e$ is depicted in Fig. \ref{fig:circuit}(d). We also show the image of $\prod_{e\in p}X_e^A$, which is useful for computational purposes, in Fig. \ref{fig:circuit}(e). Moreover, we define an operator $G_e$ in Fig. \ref{fig:circuit}(c) for each edge $e$ such that the image of $CZ^{AB}_e$ under $U$ is $CZ^{AB}_eG_e$.

The final step is to obtain plaquette operators $\hat{B}_p$ and flippers $\hat{F}_p$ acting on the Hilbert space $\mathcal{H}_A$ of $A$ qubits by restricting $U\tilde{B}_pU^\dagger$ and $UF_pU^\dagger$ respectively to the $Z_e^B=1$ subspace. $\hat{B}_p$ and $\hat{F}_p$ defined in this manner are unitary operators on $\mathcal{H}_A$ since $U\tilde{B}_pU^\dagger$ and $UF_pU^\dagger$ preserve all $Z_e^B=1$ constraints (following originally from the commutation of $\tilde{B}_p$ and $F_p$ with $C_e$). Explicitly, the resulting operators have the form
\begin{equation}
    \hat{B}_p=\hat{B}_p^0\prod_{e\in p}G_e\qquad
    \hat{F}_p=\hat{F}_p^0G_O
\end{equation}
where $\hat{F}_p^0$ and $\hat{B}_p^0$ are defined in Fig. \ref{fig:circuit}(f) and (g), and the $O$ edge is defined relative to $p$ in Fig. \ref{fig:H1}. The Hamiltonian $H_1$ on the $\mathcal{H}_A$ Hilbert space takes the form
\begin{equation}
    H_1=-\sum_p\hat{B}_p.
\end{equation}

Let us verify that the operators $\hat{B}_p$ and $\hat{F}_p$ satisfy the defining properties of a locally flippable $\mathbb{Z}_2$ separator given in Definition \ref{def:LFS}. From the first relation of Eq.~\eqref{eq:Bp2rel}, it follows that $\hat{B}_p^2=1$, satisfying property 1. Moreover, the relations
\begin{equation}
    \left[\hat{B}_p,\hat{B}_{p'}\right]=\left\{\hat{F}_p,\hat{B}_p\right\}=\left[\hat{F}_p,\hat{B}_{p'}\right]=0
    \label{eq:Fhatprop}
\end{equation}
follow from $[\tilde{B}_p,\tilde{B}_{p'}]=0$ and Eq.~\eqref{eq:Fprop}, verifying property 2. Since the $\tilde{B}_p$ operators constitute a $\mathbb{Z}_2$ separator of the $C_e=1$ subspace of the original Hilbert space, it automatically follows that the $\hat{B}_p$ operators likewise constitute a $\mathbb{Z}_2$ separator of $\mathcal{H}_A$, hence satisfying property 3. Therefore, the set of $\hat{B}_p$ and $\hat{F}_p$ operators constitute a locally flippable $\mathbb{Z}_2$ separator of the tensor product Hilbert space $\mathcal{H}_A$, thereby defining a QCA, which we call $\alpha_1$, by Theorem II.4 of Ref.~\cite{HFH18}. In the following section, we demonstrate that $H_1$ has a natural stabilizer code boundary termination that hosts chiral semion surface topological order. This provides strong evidence for the nontriviality of $\alpha_1$.

\subsection{Surface topological order} \label{sec:boundary}

In this section we extend the definition of the Hamiltonians $H_{WW}$, $H_1^{\mathbb{Z}_4}$, and $H_1$ to a lattice with boundary. The boundary Hamiltonians preserve the $\mathbb{Z}_4$ Pauli stabilizer code nature of the first two, and the $\mathbb{Z}_2$ non-Pauli stabilizer code nature of the third. For the latter two, the boundary harbors chiral semion surface topological order, whereas for $H_{WW}$, the surface hosts a semionic quasiparticle that fuses with itself into the bulk deconfined boson. Throughout, we define a truncated cubic lattice $\Lambda$ which has periodic boundary conditions in the $x$ and $y$ directions but is finite in the $z$ direction. We refer to the upper and lower boundary planes as $U$ and $L$. We demonstrate the existence of semionic surface excitations of $H_{WW}$ and $H_1^{\mathbb{Z}_4}$ by identifying semion hopping operators $\mathcal{X}_e$ associated to each edge $e$ of $\Lambda$. Whereas a string of such operators creates excitations along its length in the bulk, a string in the $U$ or $L$ plane commutes with the Hamiltonian except in the vicinity of its endpoints. The excitations at the endpoints are the deconfined surface semions. We obtain corresponding semion hopping operators $\hat{\mathcal{X}}_e$, with analogous properties, for the Hamiltonian $H_1$.

\subsubsection{Walker-Wang model}
The Hilbert space for $H_{WW}$ consists of one four-dimensional qudit on each edge of $\Lambda$. The Hamiltonian has the same form in the bulk and on the boundary:
\begin{equation}
    H_{WW}=-\sum_vA_v-\sum_p{B}_p+\text{h.c.}
\end{equation}
The bulk terms are defined in Fig. \ref{fig:H1}. The boundary terms, $A_v$ and $B_p$ for $v,p\in U,L$ are defined by a simple truncation of the bulk terms. This means that any Pauli that would act on an edge not contained in $\Lambda$ is replaced by identity.
It is also helpful to define a modified Hamiltonian which likewise has the same bulk and boundary form:
\begin{equation}
    \tilde{H}_{WW}=-\sum_vA_v-\sum_p\tilde{B}_p+\text{h.c.}
\end{equation}
The boundary terms are similarly defined by truncation of the bulk operators. Note that the terms of $H_{WW}$ and $\tilde{H}_{WW}$ remain mutually commuting products of $\mathbb{Z}_4$ Pauli operators on the lattice with boundary. We also define $C_e$ operators in the vicinity of the boundary via truncation of the bulk form, and note that the relations in Eq.~\eqref{eq:Cerel} continue to hold in the presence of a boundary.

We now introduce an operator $\mathcal{X}_e$ associated to each edge $e$ of $\Lambda$, that has the interpretation of hopping a semion along $e$ from $i$ to $j$, where $i$ ($j$) is the site adjacent to $e$ in the negative (positive) direction. We also define $\mathcal{X}_{ij}\equiv\mathcal{X}_{ji}^\dagger\equiv\mathcal{X}_e$. For edges in the bulk, $\mathcal{X}_e$ is defined in Fig. \ref{fig:H1}. For edges near the boundary, it is defined by truncating the bulk form of the operator. These operators obey the following algebraic properties:
\begin{align}
    Z_e\mathcal{X}_e=i\mathcal{X}_eZ_e,\qquad
    Z_e\mathcal{X}_{e'}=\mathcal{X}_{e'}Z_e\\
    \tilde{B}_p=\mathcal{X}_{12}\mathcal{X}_{23}\mathcal{X}_{34}\mathcal{X}_{41}Z^2_O\label{eq:BpXe}\\
    \mathcal{X}_e^2=C_e
    \label{eq:Xe2}
\end{align}
where the $O$ and $1$ through $4$ edges are defined relative to plaquette $p$ in Fig. \ref{fig:H1}. Relation Eq.~\eqref{eq:BpXe} holds in the bulk and on the lower plane $L$. However, due to truncation, for $p\in U$ we have that
\begin{equation}
    \tilde{B}_p=\mathcal{X}_{12}\mathcal{X}_{23}\mathcal{X}_{34}\mathcal{X}_{41}.
\end{equation}
Moreover, let us define the operator $\tilde{B}_p'=\tilde{B}_pA^2_1$. Then it follows that for $p\in L$
\begin{equation}
    \tilde{B}_p'=\mathcal{X}_{12}\mathcal{X}_{23}\mathcal{X}_{34}\mathcal{X}_{41}
    \left(A_1'\right)^2
\end{equation}
where $A_1'$ is a tensor product of Pauli $Z$ operators over the four edges in $L$ adjacent to vertex 1. For $p\in U$, the operator $\tilde{B}_p$ corresponds to hopping of a semion around the plaquette $p$. Likewise, for $p\in L$, the operator $\tilde{B}_p'$ corresponds to hopping of an anti-semion around $p$. (For $p\in L$, the surface anyon is a semion as viewed from below, hence an anti-semion as viewed from above). Thus, we can construct large loop operators for the surface anyons by taking products of $\tilde{B}_p$ or $\tilde{B}_p'$ operators over all plaquettes inside a given loop.
At the $U$ boundary, the anyon string operator for an oriented path $\gamma$ is given by
\begin{equation}
    W_\gamma=\prod_{e\in\gamma}\mathcal{X}_e^{(\dagger)}
    \label{eq:stringop}
\end{equation}
where the factor taken is $\mathcal{X}_e$ ($\mathcal{X}_e^\dagger$) if $e\in\gamma$ is positively (negatively) oriented. The string operator $W_\gamma$ commutes with $H_{WW}$ and $\tilde{H}_{WW}$ except in the vicinity of each endpoint, where it excites a single $A_v$ operator in addition to some $B_p$ or $\tilde{B}_p$ terms. It is clear that these excitations are fractionalized because local operators can only excite $A_v$ terms in pairs. The topological spin of the anyon is given by
\begin{equation}
    \theta=W_3W_2^\dagger W_1W_3^\dagger W_2 W_1^\dagger
    \label{eq:spin}
\end{equation}
where $W_1$, $W_2$, and $W_3$ are three open string operators that respectively move an anyon from a shared point in $U$ to three other points $r_1$, $r_2$, and $r_3$ arranged in counterclockwise order \cite{LW03,kitaev2006anyons}. Indeed we find that $\theta=i$ for the anyon on the $U$ surface, verifying that it is a semion (see Appendix~\ref{app:statistics} for details). An analogous calculation using the string operators in plane $L$ shows that the surface anyon is a semion as viewed from below.

On the other hand, a product of $\tilde{B}_p$ operators over plaquette operators comprising an open surface in the bulk does not give a loop operator, due to the factor of $Z_O^2$ in the expression in Eq.~\eqref{eq:BpXe}. This goes in hand with the fact that for $\gamma$ lying in the three-dimensional bulk, string operators of the form Eq.~\eqref{eq:stringop}  excite $B_p$ or $\tilde{B}_p$ operators along the length of the string. Physically, this represents bulk confinement of the semion excitation. However, from Eq.~\eqref{eq:Xe2} it follows that a pair of surface semions fuses into the bulk deconfined boson, which is free to move away from the surface at no energetic cost. As discussed in Sec.~\ref{sec:WW}, the hopping operator for the deconfined boson around an elementary plaquette $p$ is given by $\tilde{B}_p^2$.

\subsubsection{Short-range entangled $\mathbb{Z}_4$ stabilizer code}

We now obtain a boundary Hamiltonian for $H_1^{\mathbb{Z}_4}$ by carrying through the boson condensation procedure of Sec.~\ref{sec:cond} on the lattice with boundary. The Hilbert space is still composed of one four-dimensional qudit per edge of $\Lambda$. The Hamiltonian takes the form
\begin{equation}
    H_1^{\mathbb{Z}_4}=-\sum_{v\in L}A_v^2-\sum_p\tilde{B}_p+\text{h.c.}-\sum_eC_e
\end{equation}
where $A_v$, $\tilde{B}_p$, and $C_e$ operators near the boundary are defined by truncation of the bulk operator. The stabilizer group generated by this Hamiltonian contains all $C_e$ operators along with the all products of terms of $H_{WW}$ that commute with $C_e$.

Condensing the bulk boson destroys the bulk topological order, but leaves intact a chiral semion surface topological order. To see this, let us compute the ground state degeneracy of $H_1^{\mathbb{Z}_4}$ on the $\Lambda$ lattice. Suppose there are $N$ vertices in each $xy$ plane, and a distance of $M$ lattice spacings between the $U$ and $L$ boundary layers. Then there are $(3M+2)N$ edges and hence qudits in the lattice. For each edge the $C_e$ term gives a constraint of order 2. Apart from the $C_e$ terms, there are $(3M+2)N$ stabilizer generators, coming from $N$ vertices in plane $L$, and $(3M+1)N$ plaquettes in the lattice. The square of each of these terms is redundant, trivially in the case of $A_v^2$ and due to Eq.~\eqref{eq:Cerel} for $\tilde{B}_p$. There are also two relations between stabilizer generators due to the periodic boundary conditions:
\begin{equation}
    \prod_{p\in U}\tilde{B}_p=1,\qquad
    \prod_{v\in L}A^2_v\prod_{p\in L}\tilde{B}_p=1.
\end{equation}
Altogether, this counting implies a ground state degeneracy of 4.

For $H_1^{\mathbb{Z}_4}$, it remains the case that $\tilde{B}_p$ and $\tilde{B}_p'$, for $p\in U$ and $p\in L$ respectively, represent hopping of a surface semion around $p$. However, $C_e$ is now part of the stabilizer group, thus Eq.~\eqref{eq:Xe2} implies that the semion fuses with itself into the vacuum superselection sector. Therefore, we conclude that the boundary indeed hosts chiral semion surface topological order. When $M=0$, the model collapses to a $\mathbb{Z}_4$ Pauli stabilizer code representation of the 2D double semion topological order \cite{ellison2021pauli}.

Before moving on, we note that the bulk relation Eq.~\eqref{eq:BpXe} sheds light on the structure of the flippers $F_p$. These operators are the unique Pauli operators satisfying the commutation relations
\begin{equation}
    \left\{F_p,Z^2_O\right\}=\left[F_p,Z^2_{e\neq O}\right]=[F_p,\mathcal{X}_e]=0.
\end{equation}
These relations together with Eq.~\eqref{eq:BpXe} imply Eq.~\eqref{eq:Fprop}.

\subsubsection{Locally flippable $\mathbb{Z}_2$ separator Hamiltonian}

Finally, we extend the Hamiltonian $H_1$ to the lattice with boundary via the transformation of Sec.~\ref{sec:transformation} on the boundary Hamiltonian $H_1^{\mathbb{Z}_4}$. This amounts to first performing the qudit to two-qubit algebra automorphism Eq.~\eqref{eq:auto}, then acting with a truncated circuit $U$ such that each truncated $C_e$ term is mapped to a single Pauli $Z_e^B$, and finally projecting into the $Z_e^B=1$ subspace. The resulting Hamiltonian on $\mathcal{H}_A$ is
\begin{equation}
    H_1=-\sum_{v\in L}\hat{A}_v-\sum_p\hat{B}_p
\end{equation}
where the $\hat{B}_p$ terms in the vicinity of either boundary are truncated versions of their bulk form -- any single-qubit or two-qubit gate that would act on an edge outside $\Lambda$ is dropped from the definition of the term. Here we define
\begin{equation}
    \hat{A}_v=\prod_{e\ni v}Z_e\qquad
\end{equation}
which is a 5-body operator since $z\in L$. As is the case for the bulk Hamiltonian, all of the terms in $H_1$ in the presence of a boundary commute with one another and square to the identity, due to Eq.~\eqref{eq:Cerel} and commutation of $\tilde{B}_p$.

\begin{figure}[t]
    \centering
    \includegraphics[width=.9\textwidth]{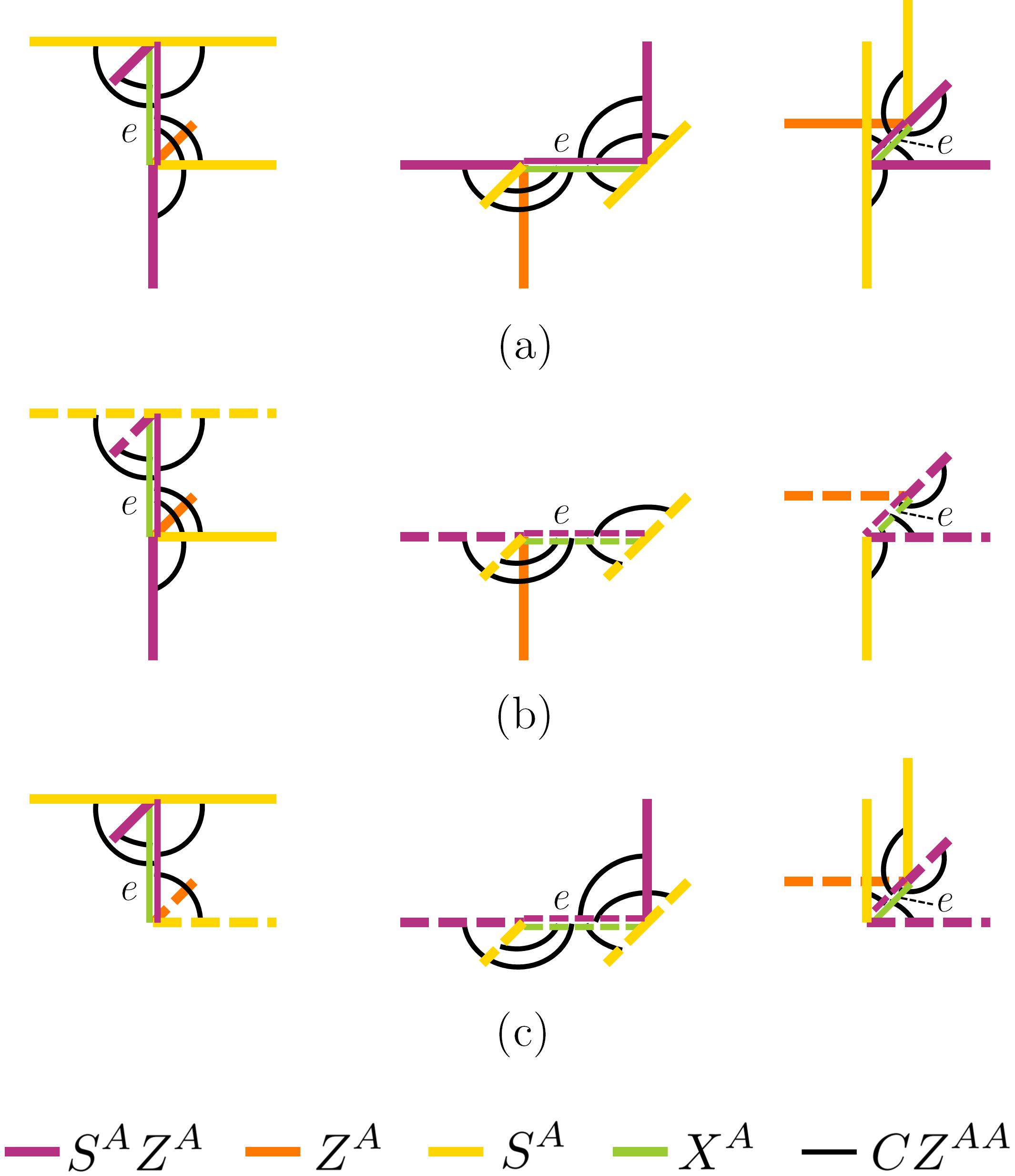}
    \caption{Semion hopping operators $\hat{\mathcal{X}}_e$ (a) in the bulk, (b) near the $U$ boundary (indicated by dashed edges), and (c) near the $L$ boundary (also indicated by dashed edges). In all cases $S$ and $CZ$ operators precede Pauli $X$ operators.}
    \label{fig:Xehat}
\end{figure}

We now introduce semion hopping operators for the Hamiltonian $H_1$. For each edge $e$ we introduce an operator $\hat{\mathcal{X}}_e$ that squares to a multiple of the identity, representing hopping of a semion in either direction along $e$. These operators, both in the bulk and near the boundary, are defined in Fig. \ref{fig:Xehat}. The semion hopping operators satisfy the following relations
\begin{equation}
    \begin{split}
    &\hat{B}_p=\hat{\mathcal{X}}_{12}\hat{\mathcal{X}}_{23}\hat{\mathcal{X}}_{34}\hat{\mathcal{X}}_{41}Z_O\hspace{.15cm}\text{if}\hspace{.15cm} p\not\in U\qquad \\
    &\hat{B}_p=\hat{\mathcal{X}}_{12}\hat{\mathcal{X}}_{23}\hat{\mathcal{X}}_{34}\hat{\mathcal{X}}_{41}\hspace{.15cm}\text{if}\hspace{.15cm} p\in U,
    \end{split}
    \label{eq:Bphat}
\end{equation}
which allow us to compute the ground state degeneracy of $H_1$.

There are a total of $(3M+2)N$ edges (and thus qubits) in the lattice. There are $(3M+1)N$ plaquettes in the lattice and $N$ vertices in the $L$ plane, and hence $(3M+2)N$ stabilizer generators in the Hamiltonian. There are two relations between these generators:
\begin{equation}
    \prod_{p\in U}\hat{B}_p=1\qquad
    \prod_{v\in L}\hat{A}_v\prod_{p\in L}\hat{B}_p=1,
\end{equation}
implying a ground state degeneracy of $4$. This is consistent with chiral semion topological order along the top boundary and anti-semion topological order along the bottom boundary (as viewed from above).

To verify this is the case, let us consider the string operators for deconfined surface excitations along the upper and lower boundaries. At the upper boundary, the anyon string operator for a path $\gamma$ is simply $\prod_{e\in\gamma}\hat{\mathcal{X}}_e$. Given the form of the string operator, we can compute the topological spin of the anyon using Eq.~\eqref{eq:spin}. Indeed we find that $\theta=i$, hence the upper surface anyon is a semion (see Appendix~\ref{app:statistics} for details).
If $\gamma$ is a contractible loop, then the semion string operator is simply the product of $\hat{B}_p$ operators inside the loop. On the other hand, at the bottom boundary the string operator for a loop is given by a product of operators of the form $\hat{B}_pA_1$ where vertex $1$ is defined in Fig.~\ref{fig:H1}. Therefore, the topological spin for anyons on the bottom layer differs by a factor of $-1$ from those on the top (due to the anticommutation of $\mathcal{X}_e$ and $Z_e$), hence they are anti-semions as viewed from above. Note that when $M=0$ (such that $A_v$ is 4-body), the model collapses to a $\mathbb{Z}_2$ non-Pauli stabilizer code representation of the double semion topological order \cite{MTE21}.

\section{Beyond chiral semion surface topological order}
\label{sec:N}

In this section we generalize the construction of $H_1$ and $\alpha_1$ in the previous section to general $n$. The generalized construction closely follows the $n=1$ version, so we omit some details for brevity. The Hamiltonian $H_n$ admits exactly-solvable boundaries hosting $U(1)_N$ surface topological order where $N=2^n$. The construction closely mirrors that of $H_1$: we begin with the Walker-Wang model based on the $\mathbb{Z}_{2N}^{(1)}$ premodular tensor category. $\mathbb{Z}_{2N}^{(1)}$ is an Abelian theory with fusion group $\mathbb{Z}_{2N}$, trivial $F$-symbols, and $R$-symbols $R^{a,b}_{a+b}=\exp{i\pi\frac{ab}{N}}$~\cite{ParsaThesis}. The generating anyon $a$ thus has topological spin $e^{i\pi/N}$, and the order two anyon $a^{N}$ is a boson with trivial braiding statistics. Therefore, the $\mathbb{Z}_{2N}^{(1)}$ Walker-Wang model has bulk $\mathbb{Z}_2$ 3D topological order, and $\mathbb{Z}_{2N}$ surface anyons $a$ such that $N$ anyons of charge $a$ fuse into the deconfined $\mathbb{Z}_2$ boson. We then condense this boson in the bulk, yielding a 3D short-range entangled state with chiral surface topological order characterized by $U(1)_N$ Chern-Simons theory. This surface order contains a single order $N$ anyon with topological spin $\theta=e^{i\pi/N}$.

To construct the generalized disentangling QCA, we again leverage the notion of a locally flippable separator and Theorem II.4 of Ref.~\cite{HFH18}. We show that the terms of $H_n$ constitute a locally flippable $\mathbb{Z}_N$ separator, hence giving rise to a QCA $\alpha_n$ which maps $H_n$ to a trivial Hamiltonian with product state eigenstates.

\begin{definition}[\cite{HFH18}]
A {locally flippable $\mathbb{Z}_N$ separator} is an indexed set of operators $B_a$ (separators) and $F_a$ (flippers), each supported in a finite radius disk, satisfying:
\begin{enumerate}
    \item{$B_a^N=1$}
    \item{$F_aB_a=e^{2\pi i/N}B_aF_a \qquad[F_a,B_b]=[B_a,B_b]=0$ for $a\neq b$.}
    \item{For an arbitrary assignment $a\mapsto\omega(a)$, where $\omega(a)$ is an $N$th root of unity, the space of states $\ket{\psi}$ such that $B_a\ket{\psi}=\omega(a)\ket{\psi}$ for all $a$ is one-dimensional.}
\end{enumerate}
\label{def:LFSN}
\end{definition}

\subsection{$\mathbb{Z}_{2N}^{(1)}$ Walker-Wang model}

\begin{figure*}[htbp]
    \centering
    \includegraphics[width=\textwidth]{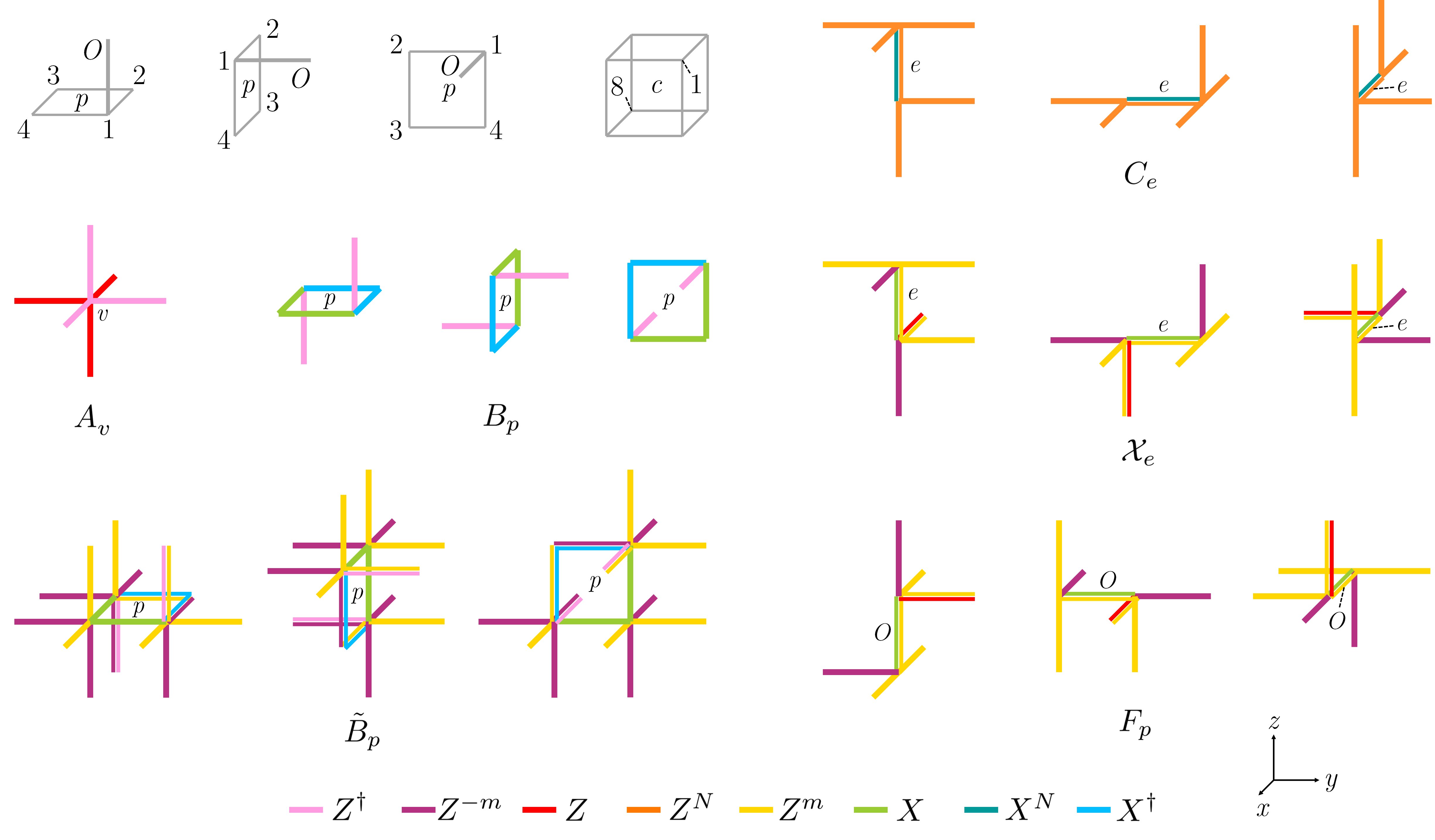}
    \caption{Vertex term $A_v$, plaquette terms $B_p,$ modified plaquette terms $\tilde{B}_p$, boson hopping operators $C_e$, anyon hopping operators $\mathcal{X}_e$, and flippers $F_p$ defined with respect to the $O$ edge of $p$, in the general $N=2^n$ case. Each operator is a tensor product of $\mathbb{Z}_{2N}$ Pauli operators over the indicated qudits, with a legend shown at the bottom. Edges with two colors represent a product of the two operators, with $Z$ type operators preceding $X$ type. All operators have an overall phase of $+1$ except the $\mathcal{X}_e$ operators which have an overall phase of $e^{-i\pi m(N-1)/2N}$. A polynomial representation of these operators is given in Appendix \ref{sec:poly}.}
    \label{fig:Hn}
\end{figure*}

The $\mathbb{Z}_{2N}^{(1)}$ Walker-Wang model is composed of $2N$-dimensional qudits on each edge of a cubic lattice, characterized by the generalized Pauli operators $Z$ and $X$ obeying the $\mathbb{Z}_{2N}$ clock and shift algebra
\begin{equation}
    Z^{2N}=X^{2N}=1\qquad ZX=e^{i\pi/N} XZ.
\end{equation}
The Hamiltonian takes the form
\begin{equation}
    H_{WW}=-\sum_vA_v-\sum_p{B}_p+\text{h.c.}
\end{equation}
whose terms are pictured in Fig.~\ref{fig:Hn}. In defining these terms, we introduce an integer
\begin{equation}
    m=\frac{1}{3}\left(\frac{3+(-1)^n}{2} 2^{n+1}-1 \right)
\end{equation}
satisfying $3m\equiv-1\mod 2N$. Note that $m=1$ in the case $n=1$.

As mentioned, $H_{WW}$ has $\mathbb{Z}_2$ bulk topological order. To verify, let us construct the topological string operators corresponding to the deconfined $\mathbb{Z}_2$ boson. We note that the $B_p$ plaquette operators satisfy the relation
\begin{equation}
    \prod_{p\in c}B_p=A_1A_8
\end{equation}
where the 1 and 8 vertices of cube $c$ are defined as in Fig. \ref{fig:Hn}. As in Sec.~\ref{sec:QCA} we have implicitly oriented each plaquette in this equation. We define modified plaquette operators
\begin{equation}
    \tilde{B}_p\equiv B_p(A_1A_3A_4)^{-m}
\end{equation}
where the 1, 3, and 4 vertices are indicated relative to $p$ in Fig.~\ref{fig:H1}. These operators satisfy the alternative property
\begin{equation}
    \prod_{p\in c}\tilde{B}_p=A_1^2
    \label{eq:BprelN}
\end{equation}
and hence
\begin{equation}
    \prod_{p\in c}\tilde{B}_p^N=1.
    \label{eq:Bp2relN}
\end{equation}
The operator $\tilde{B}_p^N$ corresponds to the motion of a deconfined $\mathbb{Z}_2$ boson around the loop. To see this, we define a set of short string operators $C_e$ for each edge $e$ of the lattice (Fig. \ref{fig:H1}), which satisfy the relations
\begin{equation}
    \tilde{B}^N_p=\prod_{e\in p}C_e\qquad C_e^2=1.
    \label{eq:CerelN}
\end{equation}
Moreover, these operators satisfy the commutation relations
\begin{equation}
    C_eA_v=\begin{cases}-A_vC_e&\text{if $v\in e$}\\+A_vC_e&\text{if $v\not\in e$}\end{cases}
    \qquad \left[C_e,\tilde{B}_p\right]=[C_e,C_{e'}]=0.
\end{equation}
A string of $C_e$ operators creates a pair of deconfined bosons at the two endpoints, characterized by the excitation of a single $A_v$ term (along with some $B_p$ terms).

\subsection{Condensing the bulk $\mathbb{Z}_2$ boson}

We now define a new $\mathbb{Z}_{2N}$ Pauli stabilizer code Hamiltonian $H_1^{\mathbb{Z}_{2N}}$ that physically represents a system obtained by condensing the deconfined $\mathbb{Z}_2$ boson excitation of $H_{WW}$. The Hilbert space is still composed of one $\mathbb{Z}_{2N}$ qudit per edge of a cubic lattice, and the Hamiltonian takes the form
\begin{equation}
    H_1^{\mathbb{Z}_{2N}}=-\sum_p\tilde{B}_p+\text{h.c.}-\sum_eC_e.
\end{equation}
The stabilizer group generated by this Hamiltonian contains all $C_e$ operators along with the all products of terms of $H_{WW}$ that commute with $C_e$. The inclusion of boson hopping operators $C_e$ in the stabilizer group implies condensation of bosons in the ground state.

The essential property of $H_n^{\mathbb{Z}_{2N}}$ is that the $\tilde{B}_p$ terms constitute a locally flippable $\mathbb{Z}_N$ separator of the $C_e=1$ constrained Hilbert space. To demonstrate this, we identify flippers $F_p$ (Fig. \ref{fig:Hn}) satisfying
\begin{equation}
    F_p\tilde{B}_p=e^{2\pi i/N}\tilde{B}_pF_p \qquad\left[F_p,\tilde{B}_{p'}\right]=[F_p,C_e]=0.
    \label{eq:FpropN}
\end{equation}
This allows us to construct a locally flippable separator of a tensor product Hilbert space by unitarily mapping the set of $C_e$ operators to single-qubit Pauli operators on unique qubit degrees of freedom. The Hamiltonian $H_n$ is obtained by projecting these qubits out from the transformed $H_n^{\mathbb{Z}_{2N}}$.

\subsection{Obtaining a locally flippable $\mathbb{Z}_N$ separator}

We now transform $H^{\mathbb{Z}_{2N}}_n$ into the $\mathbb{Z}_N$ stabilizer code Hamiltonian $H_n$ by disentangling a qubit from each edge of the lattice, leaving behind an $N$-dimensional qudit on each site.
First, we perform an operator algebra automorphism from a $2N$-dimensional qudit on each edge to a $N$-dimensional qudit tensored with a qubit:
\begin{equation}
    {\thickbar{X}}\to \hat{X}\hat{C}Z\qquad
    {\thickbar{Z}}\to X\hat{S}\qquad
    {\thickbar{X}}^N\to Z\qquad
    {\thickbar{Z}}^N\to \hat{Z}^{N/2}
    \label{eq:autoN}
\end{equation}
where ${\thickbar{Z}}$ and ${\thickbar{X}}$ represent the $\mathbb{Z}_{2N}$ Pauli operators, $Z$ and $X$ the $\mathbb{Z}_2$ Pauli operators, and $\hat{Z}$ and $\hat{X}$ the $\mathbb{Z}_N$ operators. Here, we define the operator 
\begin{equation}
    \hat{S}\equiv\hat{Z}^{1/2}=\text{diag}(1,\omega,\ldots,
\omega^{N-1})
\end{equation}
acting on the $N$-dimensional qudit with $\omega=e^{i\pi/N}$. The operator $\hat{C}Z$ acts on the total qudit-qubit Hilbert space, and is defined as
\begin{equation}
    \hat{C}Z\ket{a,b}=\begin{cases} 
      \ket{a,b} & a=0,\ldots,N-2 \\
      (-1)^{b}\ket{a,b} & a=N-1
   \end{cases}
\end{equation}
where $\ket{a,b}$ denotes a computational basis state of the qudit-qubit pair. We refer to Appendix~\ref{app: qudit automorphisms} for a group-theoretic derivation of the operator algebra automorphism in Eq.~\eqref{eq:autoN}.

Following this algebra automorphism we act with a unitary $U$, which is a translation-invariant circuit composed of generalized controlled-$X$ gates which we refer to as $\hat{C}X$ gates. The operator $\hat{C}X$ is two-body, acting on one $N$-dimensional qudit and one qubit. It acts as
\begin{equation}
    \hat{C}X\ket{a,b}=\ket{a,b+a\hspace{.1cm}\text{mod}\hspace{.1cm}2}\quad a=0,\ldots,N-1,\quad b=0,1
\end{equation}
on states, and
\begin{equation}
    Z\to \hat{Z}^{N/2}Z\qquad \hat{Z}\to \hat{Z}\qquad X\to X\qquad \hat{X}\to \hat{X}X
\end{equation}
on operators. Explicitly, $U$ has the form
\begin{equation}
    U=\prod_e\hat{C}X_{e}\prod_vU_v
\end{equation}
\begin{equation}
    \begin{split}
        U_v=\hspace{.1cm}&\hat{C}X_{23}\hat{C}X_{63}\hat{C}X_{16}\hat{C}X_{26}\hat{C}X_{56}\\
        &\hat{C}X_{12}\hat{C}X_{52}\hat{C}X_{15}\hat{C}X_{35}\hat{C}X_{45}\\
        &\hat{C}X_{31}\hat{C}X_{41}\hat{C}X_{24}\hat{C}X_{34}\hat{C}X_{64}
    \end{split}
\end{equation}
where $\hat{C}X_{ij}$ denotes a generalized controlled-$X$ gate acting on the $N$-dimensional qudit on edge $i$ and the qubit on edge $j$. The numbering of edges with respect to vertex $v$ is as depicted in Fig. \ref{fig:circuit}(a). By construction $U$ maps each $C_e$ operator to a single qubit $Z_e$, thereby disentangling each qubit from the bulk of the $\mathbb{Z}_N$ system. To obtain the $\mathbb{Z}_N$ stabilizer code Hamiltonian $H_n$ and the flippers corresponding to each of its stabilizer generators, we define operators $\hat{B}_p$ and $\hat{F}_p$ acting on the Hilbert space of $N$-dimensional qudits by restricting $U\tilde{B}_pU^\dagger$ and $UF_pU^\dagger$ to the all $Z_e=1$ subspace. The set of $\hat{B}_p$ and $\hat{F}_p$ constitute a locally flippable $\mathbb{Z}_N$ separator, therefore yielding a QCA $\alpha_n$ that disentangles the Hamiltonians $H_n$. Explicitly, according to Ref.~\cite{HFH18}, one can always find a set of locally supported commuting flippers $\hat{F}_p'$ with the properties:
\begin{align}
    \left(\hat{F}'_p\right)^N=1,\qquad\left[\hat{F}_p',\hat{F}_{p'}'\right]=0.
\end{align}
The QCA $\alpha_n$ is then defined by the following mapping of operators:
\begin{equation}
    \alpha_n\left(\hat{B}_p\right) = \hat{X}_O\qquad \alpha_n\left(\hat{F}'_p\right) = \hat{Z}_O,
\end{equation}
where the $O$ edge is defined with respect to plaquette $p$ as in Fig.~\ref{fig:Hn}. In the following section, we provide strong evidence that these QCAs are nontrivial by demonstrating that $H_n$ can be extended to a lattice with boundary, on which it harbors $U(1)_N$ topological order.

\subsection{Surface topological order}

We first extend the Hamiltonian $H_n^{\mathbb{Z}_{2N}}$ to the lattice with boundary $\Lambda$, which has periodic boundary conditions in the $x$ and $y$ directions but is finite in the $z$ direction. We refer to the upper and lower boundary planes as $U$ and $L$. The Hilbert space is composed of one four-dimensional qudit for every edge connecting sites of $\Lambda$. The Hamiltonian takes the form
\begin{equation}
    H_n^{\mathbb{Z}_{2N}}=-\sum_{v\in L}A_v^2-\sum_p\tilde{B}_p+\text{h.c.}-\sum_eC_e
\end{equation}
where $A_v$, $\tilde{B}_p$, and $C_e$ operators near the boundary are defined by truncation of the bulk Pauli operator, meaning any Pauli that would act on an edge not contained in $\Lambda$ is replaced by identity.

We now introduce an operator $\mathcal{X}_e$ associated to each edge $e$ of $\Lambda$, that has the interpretation of hopping an order $N$ anyon along $e$ from $i$ to $j$, where $i$ ($j$) is the site adjacent to $e$ in the negative (positive) direction. We also define $\mathcal{X}_{ij}\equiv\mathcal{X}_{ji}^\dagger\equiv\mathcal{X}_e$. For edges in the bulk, $\mathcal{X}_e$ is defined in Fig. \ref{fig:Hn}.  For edges near the boundary, it is defined by truncating the bulk form of the operator. These operators obey the following algebraic properties:
\begin{align}
    Z_e\mathcal{X}_e=e^{i\pi/N}\mathcal{X}_eZ_e,\qquad
    Z_e\mathcal{X}_{e'}=\mathcal{X}_{e'}Z_e\\
    \tilde{B}_p=\mathcal{X}_{12}\mathcal{X}_{23}\mathcal{X}_{34}\mathcal{X}_{41} Z^2_O\label{eq:BpXeN}\\
    \mathcal{X}_e^N=C_e
    \label{eq:XeN}
\end{align}
where the $O$ and 1 through 4 edges are indicated relative to plaquette $p$ in Fig. \ref{fig:Hn}. We have likewise defined $C_e$ operators in the vicinity of the boundary via truncation of the bulk form. Relation Eq.~\eqref{eq:BpXeN} holds in the bulk and on the lower plane $L$. However, due to truncation, for $p\in U$ we have that
\begin{equation}
    \tilde{B}_p=\mathcal{X}_{12}\mathcal{X}_{23}\mathcal{X}_{34}\mathcal{X}_{41}.
\end{equation}
Moreover, let us define the operator $\tilde{B}_p'=\tilde{B}_pA^2_1$. Then it follows that for $p\in L$
\begin{equation}
    \tilde{B}_p'=\mathcal{X}_{12}\mathcal{X}_{23}\mathcal{X}_{34}\mathcal{X}_{41}
    \left(A_1'\right)^2
\end{equation}
where $A_1'$ is the truncation of $A_1$ to the $L$ plane.
For $p\in U$, the operator $\tilde{B}_p$ corresponds to hopping of an order $N$ anyon around the plaquette $p$. Likewise, for $p\in L$, the operator $\tilde{B}_p'$ corresponds to hopping of an order $N$ anyon around $p$. These anyons have order $N$ due to the property Eq.~\eqref{eq:XeN}. We can construct large loop operators for the surface anyons by taking products of $\tilde{B}_p$ or $\tilde{B}_p'$ operators over all plaquettes inside a given loop.
At the $U$ boundary, the anyon string operator for an oriented path $\gamma$ is given by
\begin{equation}
    W_\gamma=\prod_{e\in\gamma}\mathcal{X}_e^{(\dagger)}
\end{equation}
where the factor taken is $\mathcal{X}_e$ ($\mathcal{X}_e^\dagger$) if $e\in\gamma$ is positively (negatively) oriented. The string operator $W_\gamma$ commutes with $H_{WW}$ and $\tilde{H}_{WW}$ except in the vicinity of each endpoint, where it excites a single $A_v$ operator in addition to some $B_p$ or $\tilde{B}_p$ terms. It is clear that these excitations are fractionalized because local operators can only excite $A_v$ terms in pairs. The topological spin of the anyon is given by
\begin{equation}
    \theta=W_3W_2^\dagger W_1W_3^\dagger W_2 W_1^\dagger
\end{equation}
where $W_1$, $W_2$, and $W_3$ are three open string operators that respectively move an anyon from a shared point in $U$ to three other points $r_1$, $r_2$, and $r_3$ arranged in counterclockwise order \cite{LW03,kitaev2006anyons}. We find that $\theta=e^{i\pi/N}$ for the anyon on the $U$ surface. An analogous calculation using the string operators in plane $L$ shows that the surface anyon has topological spin $\theta=e^{-i\pi/N}$ as viewed from above. Thus we conclude that the boundary harbors surface topological order characterized by $U(1)_N$ Chern-Simons theory, as claimed.

Let us compute the ground state degeneracy of this Hamiltonian. Suppose there are $N$ vertices in each $xy$ plane, and a distance of $M$ lattice spacings between the $U$ and $L$ boundary layers. Then there are $(3M+2)N$ edges and hence qudits in the lattice. For each edge the $C_e$ term gives a constraint of order 2. Apart from the $C_e$ terms, there are $(3M+2)N$ stabilizer generators, coming from $N$ vertices in plane $L$, and $(3M+1)N$ plaquettes in the lattice. The $N$th power of each of these terms is redundant, trivially in the case of $A_v^2$ and due to Eq.~\eqref{eq:CerelN} for $\tilde{B}_p$. There are also two relations between stabilizer generators due to the periodic boundary conditions:
\begin{equation}
    \prod_{p\in U}\tilde{B}_p=1\qquad
    \prod_{v\in L}A^2_v\prod_{p\in L}\tilde{B}_p=1.
\end{equation}
Altogether, this counting implies a ground state degeneracy of $N^2$, verifying the presence of $U(1)_N$ surface topological order.

To define the boundary Hamiltonian for $H_n$ with the same surface topological order, we simply carry through the transformation of the previous section on the lattice with boundary. This amounts to transforming each $C_e$ operator, in the bulk and boundary, to a single-qubit Pauli $Z$ operator.

\section{The categorical Witt group and its relation to Walker-Wang models and quantum cellular automata}
\label{sec:Witt}

In this section, we outline a set of conjectures in an attempt to formalize the growing links between the group of QCAs, Walker-Wang models, and the categorical Witt group~\cite{Haah21} (see Ref.~\cite{HaahTalk} for further discussion). First we introduce relevant background on the categorical Witt group, following the treatment in Refs.~\cite{Davydov2013a,Davydov2013b}. We then discuss the conjectures and their implications for the constructions introduced in this work.

\subsection{The categorical Witt group}

Heuristically, the categorical Witt group is the group formed by equivalence classes of anyon theories modulo boson condensation.  
Its purpose is to formalize the obstruction to a modular tensor category (MTC) containing a Lagrangian algebra, \textit{i.e.}, a set of mutually condensible bosons whose condensation trivializes the topological order.
To make the definition of Witt equivalence precise we introduce the notion of a bosonic algebra object of an MTC $\mathcal{M}$. Specifically, a bosonic algebra object $A$ corresponds to a subset of mutually bosonic anyons from $\mathcal{M}$ that contains the identity, equipped with a choice of associative fusion channels that allows them to be condensed consistently~\cite{Bais2009}.\footnote{This is formalized by the notion of a connected Et\'ale algebra in Ref.~\cite{Davydov2013a}.} Furthermore, we define $\mathcal{M}//A$ to be the MTC obtained by condensing $A$ in $\mathcal{M}$. The categorical Witt equivalence is then defined as follows:

\begin{definition}[Categorical Witt equivalence] \label{def: Witt equivalence}
MTCs $\mathcal{M}_1,\mathcal{M}_2,$ are Witt equivalent, denoted ${\mathcal{M}_1\sim \mathcal{M}_2}$, if and only if there exists bosonic algebra objects ${A_1\in \mathcal{M}_1},\, {A_2\in \mathcal{M}_2}$ such that $\mathcal{M}_1 // A_1$ and $\mathcal{M}_2 //A_2$ are braided equivalent, denoted as $\mathcal{M}_1 // A_1 \simeq \mathcal{M}_2 //A_2$.
\end{definition}

\noindent The above characterization of Witt equivalence implies that $\mathcal{M}\sim \mathcal{M}//A$ for any bosonic algebra object $A\in \mathcal{M}$, since $\mathcal{M}//1\simeq\mathcal{M}$. 
This leads to a unique representative for each Witt class.
\begin{definition}[Completely anisotropic MTC] 
An MTC $\mathcal{M}$ is completely anisotropic if the only bosonic algebra object $A\in\mathcal{M}$ is trivial, \textit{i.e.}, $A=1$ corresponding to the vacuum sector. 
\end{definition}

\begin{theorem}[\cite{Davydov2013a}]
Each Witt equivalence class contains a representative completely anisotropic MTC that is unique up to braided equivalence. 
\end{theorem}
Physically, the completely anisotropic anyon theories are those in which as many bosons as possible have been condensed. We say an MTC is Witt trivial if it belongs to the same Witt equivalence class as the trivial anyon theory, implying that it contains a Lagrangian algebra. Thus, Witt nontrivial MTCs are those that do not admit a gapped boundary.

The concept of Witt equivalence admits several equivalent characterizations, presented below for completeness. Two of the characterizations below involve the notion of the Drinfeld center $\mathcal{Z}(\mathcal{C})$ of a fusion category $\mathcal{C}$. In physical terms, $\mathcal{Z}(\mathcal{C})$ describes the emergent anyon theory of a string-net model~\cite{Levin2005} constructed from $\mathcal{C}$.
We note that, for any choice of braiding that makes $\mathcal{C}$ into a valid anyon theory (\textit{i.e.}, an MTC), then $\mathcal{Z}(\mathcal{C})\simeq \mathcal{C} \boxtimes \overline{\mathcal{C}}$, where $\overline{\mathcal{C}}$ denotes time/orientation reversal and $\boxtimes$ represents the usual stacking operation of MTCs. We refer the interested reader to Ref.~\cite{Muger2001} for more in depth discussion.

\begin{theorem}[Characterizations of Witt equivalence]
\label{prop:CWE}
For MTCs $\mathcal{M}_1,\mathcal{M}_2$, the following conditions are equivalent~\cite{Davydov2013a}:
\begin{enumerate}
    \item $\mathcal{M}_1,\mathcal{M}_2,$ are Witt equivalent, \textit{i.e.}, $\mathcal{M}_1\sim \mathcal{M}_2$.
    
    \item There exists an MTC $\mathcal{B}$ containing bosonic algebra objects $A_1,A_2\in \mathcal{B}$ such that $\mathcal{M}_1 \simeq \mathcal{B}//A_1$ and $\mathcal{M}_2 \simeq \mathcal{B}//A_2$.
    
    \item There exists fusion categories $ \mathcal{C}_1,\mathcal{C}_2,$ such that ${\mathcal{M}_1\boxtimes \mathcal{Z}(\mathcal{C}_1)\simeq \mathcal{M}_2\boxtimes \mathcal{Z}(\mathcal{C}_2)}$.
    
    \item There exists a fusion category $\mathcal{C}$ such that ${\mathcal{M}_1 \boxtimes \overline{\mathcal{M}_2} \simeq \mathcal{Z}(\mathcal{C})}$.
\end{enumerate}
\end{theorem}
The third characterization above is used to define Witt equivalence in Ref.~\cite{Davydov2013a}. 
Informally, this characterization tells us that a pair of anyon theories are Witt equivalent if they support identical emergent anyon theories after being stacked with some string-net models. 
We have started from an alternate, equivalent, definition of Witt equivalence in \ref{def: Witt equivalence} as it more closely suits our purposes. The fourth characterization has a simple physical interpretation: two MTCs are Witt equivalent if and only if there exists a gapped domain wall between the two topological orders. This can be seen via the folding trick~\cite{Kitaev2012}, whereby a domain wall between $\mathcal{M}_1$ and $\mathcal{M}_2$ is equivalent to a boundary to vacuum of $\mathcal{M}_1\boxtimes\overline{\mathcal{M}}_2$. Hence, again we see that an MTC is Witt nontrivial if and only if its boundary is non-gappable.

We now present a definition of the categorical Witt group. The Witt equivalence classes, denoted $[\mathcal{M}]$, form an  Abelian group under the stacking operation of the representative MTCs. The equivalence class of the trivial anyon theory $[\text{Vec}]$ plays the role of the identity, and the equivalence class of the braiding-reversed MTC $[\overline{\mathcal{M}}]$ provides an inverse to $[\mathcal{M}]$ corresponding to its time/orientation reversed anyon theory. The above follow from $\mathcal{M}_1\boxtimes \mathcal{M}_2\simeq \mathcal{M}_2 \boxtimes \mathcal{M}_1$, $\mathcal{M}\boxtimes \text{Vec}\simeq \mathcal{M}$ and $\mathcal{Z}(\mathcal{M})\simeq \mathcal{M} \boxtimes \overline{\mathcal{M}}$ for an MTC $\mathcal{M}$. 
Moreover, the emergent anyon theory of any string-net model is Witt equivalent to the identity element. 

\begin{definition}[Categorical Witt group]
The set of Witt equivalence classes of MTCs under stacking form an Abelian group $\mathcal{W}$ called the categorical Witt group.
\end{definition}

The categorical Witt group generalizes an older construction, which we refer to here as the classical Witt group, based on equivalence classes of metric groups.
A metric group corresponds to an Abelian group $A$ equipped with a non-degenerate quadratic form ${q:A\rightarrow U(1)}$, which physically represents an Abelian 2D topological order. 
Rather than describing the equivalence relation on metric groups directly we translate them into Abelian MTCs where the equivalence relation is inherited from the Witt equivalence relation described above.
Metric groups $(A,q)$ naturally realize Abelian (\textit{i.e.}, pointed) MTCs, denoted $\mathcal{M}(A,q)$, whose fusion is defined by the group structure $A$, and whose braiding is determined via topological spins given by the quadratic form, $\theta_a = q(a)$.
The classical Witt group $\mathcal{W}_{pt}$ of metric groups embeds into the categorical Witt group in a natural way via this mapping of metric groups to Abelian MTCs~\cite{Davydov2013a}. 

The structure of the classical Witt group is known to be
\begin{align}
    \mathcal{W}_{pt} = \bigoplus_{p \text{ prime}} \mathcal{W}_{pt}^{(p)} \, ,
\end{align}
where $\mathcal{W}_{pt}^{(p)}$ denotes the equivalence classes with $p$-group fusion rules.\footnote{That is, the anyons generate a group of the form $\bigoplus_{k \in \mathcal{K}} \mathbb{Z}_{p^{k}}$ for some set of positive integers $\mathcal{K}$.} 
These are given by: 
\begin{itemize}
    \item $\mathcal{W}_{pt}^{(2)}\cong \mathbb{Z}_8 \oplus \mathbb{Z}_2$, generated by $\mathcal{M}(\mathbb{Z}_2,q)$ and $\mathcal{M}(\mathbb{Z}_4,q')$, where $q,q'$ are any non-degenerate quadratic forms. Physically, $\mathcal{M}(\mathbb{Z}_2,q)$ corresponds to the semion or anti-semion theory. $\mathcal{M}(\mathbb{Z}_4,q')$ can be chosen to be the $U(1)_4$ anyon theory.
    \item $\mathcal{W}_{pt}^{(p)}\cong \mathbb{Z}_4$ for $p=1 \text{ mod } 4$, generated by $\mathcal{M}(\mathbb{Z}_p,q)$ with $q$ any non-degenerate quadratic form. 
    \item $\mathcal{W}_{pt}^{(p)}\cong  \mathbb{Z}_2 \oplus \mathbb{Z}_2$ for $p=3 \text{ mod } 4$, generated by $\mathcal{M}(\mathbb{Z}_p,q)$ and $\mathcal{M}(\mathbb{Z}_p,q')$, respectively, for ${q(a)=\omega^{a^2}}$ and ${q'(a)=\omega^{r a^2}}$ where $\omega$ is a primitive $p$th root of unity and $r$ is a quadratic non-residue\footnote{A number $r$ is a quadratic residue mod $p$ if it satisfies ${x^2=r \text{ mod } p}$ for some $r$, otherwise it is a quadratic non-residue.} $\mod p$. 
\end{itemize}

\subsection{QCAs, Walker-Wang models, and commuting projector Hamiltonians}

We are now in a position to state several conjectures inspired by Ref.~\cite{Haah21}, which make connections between the categorical Witt group, QCAs, Walker-Wang models, and the broader class of local commuting projector Hamiltonians with no bulk order (symmetry-breaking or topological). We define two such Hamiltonians, $H$ and $H'$, to be FDQC equivalent, denoted $H\sim H'$, if their ground states are equivalent up to the addition of ancillary qudits and the application of an FDQC.

In general, a boundary for such a Hamiltonian can be introduced by restricting the Hamiltonian to a half space and (i) including only terms fully contained therein, (ii) adding a maximal set of boundary commuting projector terms that also commute with the bulk. This results in either a gapless, extensively degenerate boundary, or a gapped boundary that may support nontrivial topological order. In the latter case, the resulting topological sectors are restricted to the vicinity of the surface and hence are expected to form an MTC $\mathcal{M}$ by the principle of braiding nondegeneracy. 
We restrict our attention to this latter case for the remainder of this section. 

The surface topological order is however, not unique, because it is always possible to stack a decoupled 2D string-net Hamiltonian based on fusion category $\mathcal{C}$ onto the boundary, resulting in surface topological order ${\mathcal{M}\boxtimes\mathcal{Z}(\mathcal{C})}$. On the other hand, we expect the Witt equivalence class of the surface topological order to be fully determined by the bulk Hamiltonian: if, to the contrary, a 3D commuting projector Hamiltonian admitted commuting projector gapped boundaries belonging to different Witt classes, we could construct a Witt nontrivial 2D commuting projector Hamiltonian by choosing Witt inequivalent upper and lower boundaries of the 3D model, then compactifying to 2D. We speculate that a similar argument forbids FDQC equivalence of commuting projector Hamiltonians with surface topological orders belonging to inequivalent Witt classes. On the other hand, in the absence of such an obstruction between a pair of commuting projector Hamiltonians, we expect that FDQC equivalence is guaranteed because their ground states are both zero-correlation length states in the trivial, short-range entangled topological phase. Summarily:
\begin{conjecture}[Condition on FDQC equivalence]
A pair of local commuting projector Hamiltonians with no long-range bulk order are FDQC equivalent if and only if their surface topological orders in the presence of a gapped boundary are Witt equivalent.
\label{conj:HEquiv}
\end{conjecture}
In other words, this conjecture states that the FDQC equivalence classes of trivially ordered commuting projector Hamiltonians are precisely labeled by the categorical Witt group.

For any MTC $\mathcal{M}$, the Walker-Wang model based on $\mathcal{M}$, denoted $H_{WW}^\mathcal{M}$, provides an example of a model with trivial bulk topological order, and a commuting projector gapped boundary condition with nontrivial surface topological order described by $\mathcal{M}$.
\begin{corollary}
$H_n\sim H_{WW}^{\mathcal{M}^{(n)}}$ where $H_n$ are the Hamiltonians introduced in this work, and $\mathcal{M}^{(n)}$ is the MTC describing $U(1)_{2^n}$ topological order. For instance $\mathcal{M}^{(1)}$ is the chiral semion MTC.
\end{corollary}
\begin{corollary}
${H_{WW}^{\mathcal{M}_1}\sim H_{WW}^{\mathcal{M}_2}}$ if and only if $\mathcal{M}_1,\mathcal{M}_2,$ are Witt equivalent, $\mathcal{M}_1\sim\mathcal{M}_2$.
\end{corollary}
This would imply the existence of an infinite family of Walker-Wang Hamiltonians based on MTCs that are short-range entangled but cannot be trivialized by FDQCs. We remark that this corollary goes beyond the usual presumption that the Walker-Wang model based on two different gauge variant formulations of the same MTC are FQDC equivalent.

A weaker statement can be obtained by taking $\mathcal{M}_2$ to be trivial: $H_{WW}^{\mathcal{M}}\sim H_{\text{triv}}$ if and only if $\mathcal{M}\simeq \mathcal{Z}(\mathcal{C})$ for some fusion category $\mathcal{C}$.
That is, the Walker-Wang model for an MTC can be disentangled by an FDQC if and only if the MTC is Witt trivial.
Disentangling circuits are known for specific examples, such as the Walker-Wang model based on toric code anyons, which is simply a cluster state~\cite{Roberts2020}. However, we are not aware of a general procedure to find disentangling circuits for an arbitrary Witt trivial MTC.

Conjecture~\ref{conj:HEquiv} identifies the Witt class as an obstruction to disentangling commuting projector Hamiltonians via FDQCs. However, recent progress has been made in constructing disentangling QCAs for specific Hamiltonians with Witt nontrivial gapped boundary conditions~\cite{HFH18,Haah21}, including the main result of this work. Thus, it is natural to speculate about the existence of QCAs that disentangle the ground states of more general Hamiltonians.
\begin{conjecture}[Existence of disentangling QCAs]
\label{conj:HQCA}
For any local commuting projector Hamiltonian with no bulk order, there exists a QCA that disentangles its ground state. A pair of disentangling QCAs are equivalent if and only if the Hamiltonians they disentangle, admit Witt equivalent surface topological orders.
\end{conjecture}
This conjecture guarantees that it is always possible to pass between FDQC equivalence classes of local commuting projector Hamiltonians via nontrivial QCA.
\begin{corollary}
For every MTC $\mathcal{M}$, there exists a QCA $Q_{WW}^{\mathcal{M}}$ that disentangles the ground state of the Walker-Wang Hamiltonian $H_{WW}^{\mathcal{M}}$.
\end{corollary}
In light of Conjecture~\ref{conj:HEquiv}, we expect $Q_{WW}^{\mathcal{M}}$ to be a nontrivial QCA precisely when $\mathcal{M}$ is Witt nontrivial. We remark that this QCA defines a locally flippable separator Hamiltonian for the Walker-Wang ground state via conjugation of the trivial Hamiltonian, $\left(Q_{WW}^{\mathcal{M}}\right)^\dagger H_\text{triv} Q_{WW}^{\mathcal{M}}$.
\begin{corollary}[Group structure of 3D QCAs]
The group of 3D QCAs contains a subgroup isomorphic to the categorical Witt group.
\end{corollary}
One could further conjecture that this subgroup is improper, or in other words the group of 3D QCAs is isomorphic to the categorical Witt group.
\begin{corollary}
The group of 3D QCAs contains a subgroup isomorphic to the classical Witt group $\mathcal{W}_{pt}$. The ground state of any commuting projector Hamiltonian with Abelian surface topological order and no bulk order can be disentangled by a QCA belonging to this subgroup.
\end{corollary}
Refs.~\cite{HFH18,Haah21} have identified a group of prime dimensional Clifford QCAs representing a subgroup of the classical Witt group containing odd prime factors $\mathcal{W}_{pt}^{(p)}$, and an additional QCA of order two corresponding to the 3-fermion MTC.
We expect our constructions to complete the realization of QCA representatives for $\mathcal{W}_{pt}$, as our construction produces QCAs corresponding to the generators of the full $\mathcal{W}_{pt}^{(2)}\cong\mathbb{Z}_8\oplus\mathbb{Z}_2$ subgroup of Witt classes containing all Abelian MTCs with 2-group fusion rules. Specifically, we conjecture that $\alpha_1$ generates the $\ZZ_8$ subgroup, and $\alpha_1 \alpha_2^{-1}$ generates the $\ZZ_2$ subgroup.
\begin{corollary} \label{cor: Abelian QCA group laws}
Let $\alpha_{3F}$ refer to the QCA of Ref.~\cite{HFH18}. Given a QCA $\alpha$, we denote the corresponding equivalence class of QCAs modulo FDQCs and translations by $[\alpha]$. Conjecture~\ref{conj:HQCA} implies the following relations on the class of novel QCAs $\alpha_n$ introduced in this work:
\begin{equation}
\begin{split}
    \big[\alpha_1\big]&=\big[\alpha_{2k-1}\big]\\
    \big[\alpha_2\big]&=\big[\alpha_{2k}\big]\\
    \big[\alpha_1^2\big]&=\big[\alpha_2^2\big]\\
    \big[\alpha_1^4\big]&=\big[\alpha_{3F}\big]\\
    \big[\alpha_1^8\big]&=\big[1\big].
\end{split}
\end{equation}
\end{corollary}
In Appendix~\ref{app:3fsquare}, we describe a QCA $\tilde\alpha_{3F}$ in the class $[\alpha_{3F}]$, with the property that $\tilde\alpha_{3F}^2=1$, \textit{i.e.}, the QCA squares exactly to the identity.

\section{Discussion}
\label{sec:discussion}

In this work, we have introduced an exactly-solvable Hamiltonian $H_1$ with a short-range entangled bulk that supports a boundary with chiral semion surface topological order. 
The key feature differentiating our Hamiltonian from the chiral semion Walker-Wang model is that it constitutes a locally flippable separator. We have leveraged this property to define a new three dimensional QCA $\alpha_1$ that disentangles our Hamiltonian, mapping it to a sum of Pauli $Z$ operators. 
We conjecture that this QCA is nontrivial on the grounds that there would otherwise exist a standalone commuting projector Hamiltonian realizing chiral semion topological order in two dimensions. 
We have also generalized our Hamiltonian to a family of locally flippable separators $H_n$ and corresponding QCAs $\alpha_n$ based on the $U(1)_{2^n}$ Chern-Simons theories. 

Furthermore, we conjecture that the QCAs presented in this work generate a $\mathbb{Z}_8 \oplus \mathbb{Z}_2$ group. We speculate that, together with the QCAs constructed in Ref.~\cite{Haah21}, this covers all equivalence classes of QCAs that disentangle locally flippable separators with Abelian boundary topological order. This agrees with the conjecture that the group of such QCAs is isomorphic to the classical Witt group \cite{Haah21}. An important future direction is thus to rigorously verify the $\mathbb{Z}_8\oplus\mathbb{Z}_2$ group structure of our QCAs. This is a difficult calculation due to the non-Clifford nature of the QCAs. In Sec.~\ref{sec:Witt}, we also raised the challenge of assessing, more generally, whether there is
% rigorously establishing 
an isomorphism between the group of QCAs and the categorical Witt group, which includes MTCs with non-Abelian anyons. An important step in this direction is to identify an example of a QCA that disentangles a state with non-Abelian surface topological order.

Moving forwards, it would be interesting to study the classification of three-dimensional QCAs with various qualifiers on the set of QCAs. One could restrict the classification to the set of Clifford QCAs, as considered in Ref.~\cite{Haah21} for prime-dimensional qudits. It may also be enlightening to consider Clifford QCAs on composite-dimensional qudits. While the QCAs constructed in this work are non-Clifford, it is possible that they admit representations in this class. 

Another question pertains to the classification of quasi-locality-preserving QCAs, as opposed to the strictly locality-preserving QCAs discussed in this work. It is natural to consider two quasi-locality preserving QCAs to be equivalent if they differ by finite time evolution under a local Hamiltonian. It is unclear, but an important open question, whether the QCAs introduced in this work are nontrivial in this broader sense. We note that there has been recent progress on the classification of one-dimensional quasi-locality preserving QCAs in Ref.~\cite{RWW20}.

Yet another direction for future work is the classification of three-dimensional QCAs on fermionic systems. On one hand, we expect that certain QCAs become trivial in the presence of fermionic ancillas. This is because the class of fermionic local commuting projector Hamiltonians captures a wider range of 2D topological orders than its bosonic counterpart. For example, the phases of Kitaev's sixteen-fold way can be modeled by fermionic commuting projector Hamiltonians after stacking with chiral $p+ip$ superconducting states~\cite{Ware2016,Tarantino2016}. Thus, we expect the disentangling QCAs for the three-fermion Walker-Wang model~\cite{HFH18}, as well as our QCA $\alpha_2$, which disentangles the Hamiltonian $H_2$ with $U(1)_4$ surface topological order, to be trivial for fermionic systems.\footnote{$U(1)_4$ topological order is the same as that of the $\nu=2$ Kitaev sixteen-fold way state.} On the other hand, fermionic degrees of freedom may allow for nontrivial QCAs that are not possible on qudits. We expect that the connection between the classification of qudit QCAs and the categorical Witt group, discussed in Sec.~\ref{sec:Witt}, extends to an analogous connection between the classification of fermionic QCAs and the super-Witt group, \textit{i.e.}, the generalization of the categorical Witt group to 2D fermionic topological orders.

Finally, it would be worthwhile to find tensor network representations of the known nontrivial 3D QCAs~\cite{Piroli2020}. It would also be interesting to study their circuit complexity, which we expect to be of linear depth. Such methods could allow for the construction of a wider class of nontrivial QCAs.

\vspace{0.2in}
\noindent{\it Acknowledgements} --- W.S. thanks Shu-Heng Shao, Xie Chen, and John McGreevy for inspiring discussions. A.D. thanks Jeongwan Haah and Matthew Hastings for useful feedback. T.D.E. thanks Lukasz Fidkowski for valuable conversations about Ref.~\cite{HFH18}. This work was supported by the JQI fellowship at the University of Maryland (YC), the Simons Foundation through the collaboration on Ultra-Quantum Matter (651438, AD; 651444, WS), NSERC (NT), the It from Qubit collaboration (DJW), and by the Institute for Quantum Information and Matter, an NSF Physics Frontiers Center (PHY-1733907, AD, WS).

% \bibliography{bib}
% \bibliographystyle{apsrev4-2}


\appendix

\section{Coupled layers construction of $H_{WW}$} \label{app: coupledlayer}

In this appendix, we describe a simple coupled layers interpretation of $H_{WW}$ \cite{WangSenthil13,JQ14}. The recipe is to start with a stack of 2D $\mathbb{Z}_4$ toric codes in the $z$ direction, then to couple adjacent layers by condensing all anyons of the form $e_im_ie_{i+1}m_{i+1}^3$. Here the subscript indexes the layer number. As a result, all anyons in a given layer except for $e_i^2m_i^2$ are confined in the bulk due to nontrivial braiding statistics with at least one of these condensed anyons. However, pairs of these bosons, $e_i^2m_i^2e_{i+1}^2m_{i+1}^2$, also become condensed, thus giving rise to a single species of fully deconfined 3D boson with $\mathbb{Z}_2$ fusion rules. However, at the upper boundary layer $U$, the $e_U m_U$ semion survives as a deconfined surface anyon. Likewise, the $e_Lm_L^3$ anti-semion survives as a surface anyon at the lower boundary layer $L$.

$H_{WW}$ microscopically realizes the condensed phase in the following sense. Begin with a stack of $\mathbb{Z}_4$ toric codes composed of four-dimensional qudits on $x$ and $y$ edges of a cubic lattice, along with ancillary qudits on each $z$ edge $e_z$ of the lattice stabilized by the Hamiltonian $H_0=X+X^\dagger$. The total Hamiltonian is
\begin{equation}
    H=\sum_{e_z} H_0-\sum_l\left(\sum_vA_v^l+\sum_{p_z}{\hat{B}}_{p_z}^l\right)+\text{h.c.}
\end{equation}
where $A_v^l$ and $\hat{B}_{p_z}^l$ are the vertex and plaquette terms of the $\mathbb{Z}_4$ toric code on layer $l$. (Here, $p_z$ indexes plaquettes normal to the $z$ direction). Then, we may interpret the plaquette terms ${B}_{p_x}$ and ${B}_{p_y}$ of $H_{WW}$ as hopping operators for the $e_im_ie_{i+1}m_{i+1}^3$ anyons. The remaining terms in $H_{WW}$ generate the set of all products of terms in $H$ that commute with all ${B}_{p_x}$ and ${B}_{p_y}$. Turning on the hopping operators ${B}_{p_x}$ and ${B}_{p_y}$ for the $e_im_ie_{i+1}m_{i+1}^3$ anyons with coefficient $J$ gives the coupled layers Hamiltonian
\begin{align}
        H&=\sum_{e_z} H_0-\sum_l\left(\sum_vA_v^l+\sum_{p_z}{\hat{B}}_{p_z}^l\right)\\
        &~~-J\left(\sum_{p_x}B_{p_x}+\sum_{p_y}B_{p_y}\right)+\text{h.c.}
\end{align}
Taking the strong coupling limit $J\rightarrow\infty$ and applying degenerate perturbation theory in $1/J$ yields an effective Hamiltonian whose ground state is equivalent to that of $H_{WW}$.

\section{Polynomial representation of Pauli operators}
\label{sec:poly}

In this appendix, we express the Pauli operators $\mathcal X_e$, $\tilde{B}_p$, and $F_p$ of Fig.~\ref{fig:H1} as vectors over the Laurent polynomial ring $\mathbb{Z}_4\left[x,y,z,\frac{1}{x},\frac{1}{y},\frac{1}{z}\right]$, which is useful for verifying their commutation relations~\cite{Haah13}. These calculations are given in the supplementary Mathematica file. We identify the unit cell as the three qubits on edges linking the origin and $\hat{x}$, $\hat{y}$, and $\hat{z}$ respectively. We list only the operators associated with the edge $e_x$ centered at $\left(\frac{1}{2},0,0\right)$ and plaquette $p_x$ centered at $\left(0,\frac{1}{2},\frac{1}{2}\right)$. The operators associated with $e_y$, $e_z$, $p_y$, and $p_z$ can be obtained by cyclic permutation $x\to y\to z\to x$ of the matrix entries and $1\to2\to3\to1$ of the row indices.
\begin{align}
    \begin{gathered}
      \mathcal X_{e_x} =
    \begin{pmatrix}
    1\\0\\0\\
    1-\frac{1}{x}\\-\left(x+\frac{2}{y}\right)\\x+\frac{x}{z}+1
    \end{pmatrix}\nonumber\\
    \tilde{B}_{p_x}=
    \begin{pmatrix}
    0\\1-z\\y-1\\y+1-\frac{yz+y+2}{x}\\
    yz+y-\frac{1}{y}-z\\yz+1-\frac{y+1}{z}
    \end{pmatrix}\qquad
    F_{p_x}=\begin{pmatrix}yz\\0\\0\\yz-xyz\\xz+yz+z\\-(2xyz+y)\end{pmatrix}.
    \end{gathered}
\end{align}
These matrices satisfy the relations,
\begin{equation}
    \tilde{B}_{p_i}^\dagger\Omega\tilde{B}_{p_j}= 2\tilde{B}_{p_i}^\dagger\Omega \mathcal{X}_{e_j}=F_{p_i}^\dagger\Omega \mathcal{X}_{e_j}=0\qquad
    \tilde{B}_{p_i}^\dagger\Omega F_{p_j}=2\delta_{ij}
\end{equation}
where $\Omega$ is the symplectic form
\begin{equation}
    \Omega=\begin{pmatrix}0&I_3\\-I_3&0\end{pmatrix}
\end{equation}
and $^\dagger$ represents transposition combined with spatial inversion.

Next, we write the corresponding Pauli operators for the general $N=2^n$ case of Fig.~\ref{fig:Hn} as vectors over the Laurent polynomial ring $\mathbb{Z}_{2N}\left[x,y,z,\frac{1}{x},\frac{1}{y},\frac{1}{z}\right]$.
\begin{align}
    \begin{gathered}
      \mathcal X_{e_x} =
    \begin{pmatrix}
    1\\0\\0\\
    m\left(1-\frac{1}{x}\right)\\-m\left(x+\frac{2}{y}\right)\\m\left(x+\frac{x}{z}+1\right)
    \end{pmatrix}\\
    \tilde{B}_{p_x}=
    \begin{pmatrix}
    0\\1-z\\y-1\\m(yz+y+1)-yz-m\left(\frac{yz+y+1}{x}\right)-\frac{1}{x}\\
    m\left(yz+y-\frac{1}{y}-z\right)\\m\left(yz+1-\frac{y+1}{z}\right)
    \end{pmatrix}\\
        F_{p_x}=\begin{pmatrix}yz\\0\\0\\m(yz-xyz)\\m(xz+yz+z)\\-m(2xyz+y)\end{pmatrix}.
    \end{gathered}
\end{align}
These matrices satisfy the relations
\begin{equation}
    \tilde{B}_{p_i}^\dagger\Omega\tilde{B}_{p_j}= N\tilde{B}_{p_i}^\dagger\Omega \mathcal{X}_{e_j}=F_{p_i}^\dagger\Omega \mathcal{X}_{e_j}=0\qquad
    \tilde{B}_{p_i}^\dagger\Omega F_{p_j}=2\delta_{ij}
\end{equation}
via the property $3m=-1$.

\section{Operator algebra automorphisms from group extensions}
\label{app: qudit automorphisms}

In Sec.~\ref{sec:N}, we used an operator algebra automorphism to map a $2N$-dimensional qudit to a $N$-dimensional qudit and a qubit. Here, we show that this transformation can be derived from a nontrivial central extension of $\ZZ_N$ by $\ZZ_2$. We also show that an operator algebra automorphism is defined by the central extension of $\ZZ_2$ by $\ZZ_N$. Similar mappings were observed in Appendix A of Ref.~\cite{Ellison2019disentangling} and Appendix A of Ref.~\cite{Tantivasadakarn2021nonAbelian}.

As a warm-up, we start with the case $N=2$. This corresponds to the operator algebra automorphism used in Sec.~\ref{sec:transformation}. The nontrivial extension of $\ZZ_2$ by $\ZZ_2$ can be organized into the short exact sequence:
\begin{align}
0 \to \ZZ_2 \to \ZZ_4 \to \ZZ_2 \to 0.
\end{align}
This corresponds to a $2$-cocycle $[\lambda]$ belonging to $H^2[\ZZ_2,\ZZ_2]$. $[\lambda]$ may be represented by the function $\lambda: \ZZ_2 \times \ZZ_2 \to \ZZ_2$ defined by:
\begin{align}
\lambda(a_1,a_2) = \begin{cases}
1 & \text{if }a_1+a_2=2 \\
0& \text{otherwise}.
\end{cases}
\end{align}
The elements of $\ZZ_4$ can then be written as pairs $(a,b)$ in $\ZZ_2 \times \ZZ_2$ as in Table~\ref{table: qubitqubitmapping} with the group law determined by $\lambda$ as:
\begin{align}
(a_1,b_1) + (a_2,b_2) = (a_1+a_2, b_1+b_2 +\lambda(a_1,a_2)).
\end{align}

To construct an operator algebra automorphism using the group extension above, we consider a four-dimensional qudit with basis states labeled by $n \in \ZZ_4$ and a pair of qubits with basis states labeled by $(a,b) \in \ZZ_2 \times \ZZ_2$. Table~\ref{table: qubitqubitmapping} then gives a mapping between the four-dimensional qudit and the pair of qubits:
\begin{align}
|n\rangle \leftrightarrow |a,b \rangle.
\end{align}

This mapping of states determines an operator algebra automorphism, as described below. To see this, we consider the action of the the four-dimensional qudit Pauli operators $\thickbar{X}$ and $\thickbar{Z}$ on $|n\rangle$. For $\thickbar{X}$, we have:
\begin{align}
\thickbar{X}|n\rangle = |n+1 \rangle,
\end{align}
with addition modulo $4$. In terms of the $(a,b)$ labels, $\thickbar{X}$ implements the transformation:
\begin{eqs}
\thickbar{X} : |a,b \rangle \mapsto &|(a,b)+(1,0) \rangle \\
=&|a+1,b+\lambda(a,1)\rangle.
\end{eqs}
$\lambda(a,1)$ is $1$ if $a=1$. Therefore, the action of $\thickbar{X}$ is:
\begin{align}
\thickbar{X} : |a,b\rangle \mapsto X^A CX^{AB} |a,b \rangle,
\end{align}
where $A$ and $B$ label the sites of the qubits. As for Pauli $\thickbar{Z}$, we have:
\begin{align}
\thickbar{Z} |n\rangle = i^n |n\rangle,
\end{align}
which in terms of the states labeled by $(a,b)$ is:
\begin{align}
\thickbar{Z} : |a,b \rangle \mapsto S^A Z^B |a,b \rangle.
\end{align}
This can be checked from Table~\ref{table: qubitqubitmapping}. The operator algebra automorphism is thus:
\begin{eqs}
\thickbar{X} \leftrightarrow X^A CX^{AB}, \qquad \thickbar{Z} \leftrightarrow S^A Z^B.
\end{eqs}
Conjugating the $B$ site qubit by a Hadamard, we arrive at the mapping in Eq.~\eqref{eq:auto}.

\begin{table}[t]
    \small
    \centering
        \begin{tabular}{ c| c} 
            $n \in \ZZ_4$ & $\, (a,b) \in \ZZ_2 \times \ZZ_2 \,$\\
    %\specialrule{.2em}{.2em}{.2em}        0 & (0,0) \\
    \hline
             1 & (1,0) \\
             2 & (0,1) \\
             3 & (1,1) 
        \end{tabular}
\caption{The elements of $\ZZ_4$ labeled by elements of $\ZZ_2 \times \ZZ_2$.}
\label{table: qubitqubitmapping}
\end{table}

Next, we construct an operator algebra automorphism from a $2N$-dimensional qudit to an $N$-dimensional qudit and a qubit, using the group extension:
\begin{align}
0 \to \ZZ_2 \to \ZZ_{2N} \to \ZZ_N \to 0.
\end{align}
This corresponds to a $2$-cocycle $[\lambda] \in H^2[\ZZ_N,\ZZ_2]$. $[\lambda]$ may be represented by the function $\lambda : \ZZ_N \times \ZZ_N \to \ZZ_2$, given by:
\begin{align}
\lambda(a_1,a_2) = \begin{cases}
1 & \text{if }a_1+a_2 \geq N \\
0& \text{otherwise}.
\end{cases}
\end{align}
This can be written as:
\begin{align}
\lambda(a_1,a_2) = a_1 + a_2 - [a_1 + a_2]_N,
\end{align}
where $[ \,\, \cdot \,\,]_N$ denotes addition modulo $N$. We may then write elements of $\ZZ_{2N}$ as pairs in $\ZZ_N \times \ZZ_2$ according to Table~\ref{table: quditquditmapping}. The group law between $(a_1,b_1),(a_2,b_2) \in \ZZ_N \times \ZZ_2$ is determined by $\lambda$ as:
\begin{eqs}
(a_1,b_1) + (a_2,b_2) = (a_1 + a_2, b_1 + b_2 + \lambda(a_1,a_2)).
\end{eqs}

To write down an operator algebra automorphism from the mapping in Table~\ref{table: quditquditmapping}, we label the quantum states of a $2N$-dimensional qudit by elements $n \in \ZZ_{2N}$ and label the states of a $N$-dimensional qudit and a qubit by the pair $(a,b) \in \ZZ_N \times \ZZ_2$. The mapping of operators is then determined by the association of states:
\begin{align}
|n \rangle \leftrightarrow |a,b \rangle,
\end{align}
according to Table~\ref{table: quditquditmapping}.

\begin{table}
    \small
    \centering
        \begin{tabular}{ c|c } 
            $n \in \ZZ_{2N}$ &  $\, (a,b) \in \ZZ_N \times \ZZ_2 \,$\\ \hline% \specialrule{.2em}{.2em}{.2em}  
            0 & (0,0)\\
            1 & (1,0)\\
            2 & (2,0)\\
            $\vdots$ & $\vdots$\\
            $N-1$ & $(N-1,0)$\\
            $N$ & $(0,1)$\\
            $N+1$ & $(1,1)$\\
            $\vdots$ & $\vdots$\\ 
            $2N-2$ & $(N-2, 1)$\\
            $2N-1$ & $(N-1,1)$
        \end{tabular}
\caption{The elements of $\ZZ_{2N}$ labeled by elements of $\ZZ_N \times \ZZ_2$.}
\label{table: quditquditmapping}
\end{table}

We find the corresponding mapping of operators by considering the action of the $2N$-dimensional Pauli operators $\thickbar{X}$ and $\thickbar{Z}$ on $|n\rangle$. The action of $\thickbar{X}$ on $|n\rangle$ is
\begin{align}
\thickbar{X}|n\rangle = |n+1 \rangle,
\end{align}
with addition modulo $2N$. Expressed in terms of the states labeled by $(a,b)$, $\thickbar{X}$ enacts the mapping:
\begin{eqs}
\thickbar{X}: |a,b \rangle \mapsto &|(a,b)+(1,0)\rangle \\
=&|a+1,b+\lambda(a,1) \rangle.
\end{eqs}
$\lambda(a,1)$ is $1$ if and only if $a$ is $N-1$. We define $\hat{C}X$ to be the operator that acts as $X$ on the qubit if the state of the $N$-dimensional qudit is $N-1$ and acts as the identity otherwise:
\begin{align}
    \hat{C}X |a,b \rangle = \begin{cases}
|a,b+1 \rangle & \text{if } a = N-1 \\
|a,b \rangle & \text{otherwise}.
\end{cases}
\end{align}
With this, the action of $\thickbar{X}$ is
\begin{align}
\thickbar{X}: |a,b \rangle \mapsto \hat{X} \hat{C}X |a,b \rangle,
\end{align} 
with $\hat{X}$ denoting the Pauli $X$ operator on the $N$-dimensional qudit.
As for the action of $\thickbar{Z}$, we have:
\begin{align}
\thickbar{Z}|n \rangle = \omega^n |n\rangle,
\end{align}
where $\omega = \text{exp}(2\pi i/2N)$. The action of $\thickbar{Z}$ in terms of the states labeled by $(a,b)$ is
\begin{align}
\thickbar{Z}: |a,b \rangle \mapsto \hat{Z}^{1/2} Z |a,b \rangle.
\end{align}
Here, $\hat{Z}$ is the Pauli $Z$ operator on the $N$-dimensional qudit.
Note that the eigenvalues of $\hat{Z}^{1/2}$ are multiples of $\omega$, and $Z$ gives the necessary $-1$ eigenvalue, when $a$ cycles past $N-1$. We now have the operator algebra automorphism:
\begin{align}
\thickbar{X} \leftrightarrow \hat{X}\hat{C}X, \quad \thickbar{Z} \leftrightarrow \hat{Z}^{1/2} Z.
\end{align}
By conjugating the $B$ site by a Hadamard, we obtain the mapping in Eq.~\eqref{eq:autoN}.

\begin{table}
    \small
    \centering
        \begin{tabular}{ c|c } 
            $n \in \ZZ_{2N}$ & $\, (a,b) \in \ZZ_N \times \ZZ_2 \,$ \\ \hline%\specialrule{.2em}{.2em}{.2em}
            0 & (0,0)\\
            1 & (0,1)\\
            2 & (1,0)\\
            3 & (1,1)\\
            $\vdots$ & $\cdots$\\
            $2N-2$ & $(N-1,0)$\\
            $2N-1$ & $(N-1, 1)$\\
        \end{tabular}
\caption{An alternative labeling of the elements of $\ZZ_{2N}$ by elements of $\ZZ_N \times \ZZ_2$.}
\label{table: quditquditmapping2}
\end{table}

For illustrative purposes, we show that there is an alternative operator algebra automorphism from a $2N$-dimensional qudit to a $N$-dimensional qudit and a qubit, which arises from an extension of $\ZZ_2$ by $\ZZ_N$:
\begin{align}
    0 \to \ZZ_N \to \ZZ_{2N} \to \ZZ_2 \to 0.
\end{align}
In this case, the extension is characterized by $[\lambda]\in H^2[\ZZ_2,\ZZ_N]$, which may be represented by the function $\lambda: \ZZ_2 \times \ZZ_2 \to \ZZ_N$, defined by:
\begin{align}
\lambda(a_1,a_2) = \begin{cases}
1 & \text{if }a_1+a_2 = 2 \\
0& \text{otherwise}.
\end{cases}
\end{align}
The elements of $\ZZ_{2N}$ can be written as pairs $(a,b) \in \ZZ_N \times \ZZ_2$ according to the Table~\ref{table: quditquditmapping2}. The group law is
\begin{eqs}
(a_1,b_1) + (a_2,b_2) = (a_1 + a_2 + \lambda(b_1,b_2), b_1 + b_2).
\end{eqs}

We next define a basis for a $2N$-dimensional qudit with basis states labeled by $n \in \ZZ_{2N}$ and define a basis for an $N$-dimensional qudit and a qubit labeled by $(a,b)$. We then map between the two bases according to Table~\ref{table: quditquditmapping2}:
\begin{align}
|n\rangle \leftrightarrow |a,b \rangle.
\end{align}

Again, we identify a mapping of operators by considering the actions of $\thickbar{X}$ and $\thickbar{Z}$ on $|n \rangle$. The action of $\thickbar{X}$ is
\begin{align}
\thickbar{X}|n \rangle = |n+1 \rangle,
\end{align}
which, in terms of the basis for the $N$-dimensional qudit and qubit, gives the mapping:
\begin{eqs}
\thickbar{X}: |a,b \rangle \mapsto &|(a,b)+(0,1) \rangle \\
=&|a+ \lambda(b,1),b+1 \rangle \\
=&XC\hat{X}|a,b \rangle,
\end{eqs}
where $C\hat{X}$ is the CNOT gate with the qubit as the control.
The action of $\thickbar{Z}$ is
\begin{align}
\thickbar{Z} |n\rangle = \omega^n |n \rangle,
\end{align}
giving the mapping:
\begin{eqs}
\thickbar{Z}: |a,b \rangle \mapsto Z^{1/N}\hat{Z} |a,b \rangle.
\end{eqs}
This means that the operator algebra automorphism is 
\begin{align}
\thickbar{X} \leftrightarrow XC\hat{X}, \quad \thickbar{Z} \leftrightarrow Z^{1/N}\hat{Z}.
\end{align}

\section{Calculation of surface anyon statistics}
\label{app:statistics}

In this appendix we explicitly calculate the topological spin of the upper surface anyons of the models $H_1^{\mathbb{Z}_4}$ and $H_1$ defined on the lattice with boundary $\Lambda$ in Sec.~\ref{sec:boundary}. For $H_1^{\mathbb{Z}_4}$, the string operator is given in Eq.~\eqref{eq:stringop}. For edges $e$ and $e'$ that do not share a vertex, it holds that
\begin{equation}
    \left[\mathcal{X}_e,\mathcal{X}_{e'}\right]=0.
\end{equation}
Therefore, to compute the topological spin, it suffices to use the minimal, \textit{i.e.} unit length string operators. In particular, we apply Eq.~\eqref{eq:spin} with
\begin{equation}
\begin{split}
    W_1&=\mathcal{X}_1\\
    W_2&=\mathcal{X}_2\\
    W_3&=\mathcal{X}^\dagger_3
\end{split}
\end{equation}
where edges 1, 2, and 3 respectively join the origin with $(1,0,0)$, $(0,1,0)$, and $(-1,0,0)$. We find that
\begin{equation}
    \begin{split}
        \theta&=Z_3^\dagger X_3^\dagger Z_2Z_2^\dagger X_2^\dagger Z_1^\dagger X_1Z_1Z_3^\dagger X_3Z_3Z_2^\dagger X_2Z_2Z_1Z_1^\dagger X_1^\dagger Z_3\\
        &=Z_1^\dagger X_1Z_1X_1^\dagger X_2^\dagger Z_2^\dagger X_2Z_2Z_3^\dagger X_3^\dagger Z_3^\dagger X_3Z_3  Z_3\\
        &=-iX_2^\dagger Z_2^\dagger X_2Z_2Z_3^\dagger X_3^\dagger Z_3^\dagger X_3Z_3  Z_3\\
        &=-Z_3^\dagger X_3^\dagger Z_3^\dagger X_3Z_3  Z_3\\
        &=i,
    \end{split}
\end{equation}
hence the upper surface anyon is a semion. We give a similar calculation in the supplementary Mathematica file.

Next, we consider $H_1$. In this case the string operators are given by products of $\hat{\mathcal{X}}_e$ operators. Again we apply Eq.~\eqref{eq:spin} with
\begin{equation}
\begin{split}
    W_1&=\hat{\mathcal{X}}_1\\
    W_2&=\hat{\mathcal{X}}_2\\
    W_3&=\hat{\mathcal{X}}^\dagger_3.
\end{split}
\end{equation}
To facilitate the calculation, we define modified versions of these operators that have been truncated to the 1, 2, and 3 edges: 
\begin{equation}
    \begin{split}
       \tilde{W}_1&=X_1S_1^\dagger S_3^\dagger CZ_{13}\qquad \tilde{W}_1^\dagger=S_1S_3 CZ_{13}X_1\\
        \tilde{W}_2&=X_2S_2^\dagger S_1^\dagger CZ_{12}\qquad \tilde{W}_2^\dagger=S_2S_1 CZ_{12}X_2\\
        \tilde{W}_3&=S_3S_2^\dagger CZ_{23}X_3\qquad \tilde{W}_3^\dagger=X_3S_3^\dagger S_2 CZ_{23},
    \end{split}
\end{equation}
noting that
\begin{equation}
    W_3W_2^\dagger W_1W_3^\dagger W_2W_1^\dagger=\tilde{W}_3\tilde{W}_2^\dagger \tilde{W}_1\tilde{W}_3^\dagger \tilde{W}_2\tilde{W}_1^\dagger.
\end{equation}
The topological spin can then be computed using the relations
\begin{equation}
    SX=iXSZ\qquad CZ_{12}X_1=X_1Z_2CZ_{12}.
\end{equation}
Explicitly,
\vspace{5mm}
\begin{widetext}
% \onecolumngrid
\begin{equation}
\begin{split}
    \theta&= S_3S_2^\dagger CZ_{23}X_3 S_2S_1CZ_{12}X_2 X_1\mathbf{S_1^\dagger }S_3^\dagger \mathbf{CZ_{13}} X_3S_3^\dagger S_2CZ_{23}  X_2S_2^\dagger \mathbf{S_1^\dagger CZ_{12}}\mathbf{S_1} S_3\mathbf{CZ_{13}}X_1\\
    &=-iS_3S_2^\dagger CZ_{23}X_3 S_2S_1 CZ_{12}X_2S_1S_3CZ_{13} X_3S_3^\dagger S_2CZ_{23} X_2S_2S_1CZ_{12}S_1^\dagger S_3^\dagger CZ_{13}\\
    &=-iZ_1S_3S_2^\dagger CZ_{23}X_3 S_2CZ_{12}X_2 S_3CZ_{13} X_3S_3^\dagger \mathbf{S_2 CZ_{23}} X_2S_2CZ_{12}S_3^\dagger CZ_{13}\\
    &=Z_1S_3S_2^\dagger CZ_{23}X_3 S_2CZ_{12} S_3CZ_{13} X_3S_3S_2^\dagger CZ_{23}S_2 CZ_{12} S_3^\dagger CZ_{13}\\
    &=Z_1S_3CZ_{23}X_3 \mathbf{S_3CZ_{13}} X_3CZ_{23} CZ_{13}\\
    &=iZ_1S_3CZ_{23}S_3^\dagger Z_1CZ_{13} CZ_{23} CZ_{13}\\&=i.
\end{split}
\end{equation}
\end{widetext}
In the first (third, fifth) line, the bold operators are those lying between the two $X_1$ ($X_2$, $X_3$) operators that fail to commute with $X_1$ ($X_2$, $X_3$).

\section{A 3-fermion QCA that squares to the identity} \label{app:3fsquare}

\begin{figure*}[ht!]
    \centering
    \includegraphics[width=.7\textwidth, trim={7.5cm 5cm 10cm 5cm},clip]{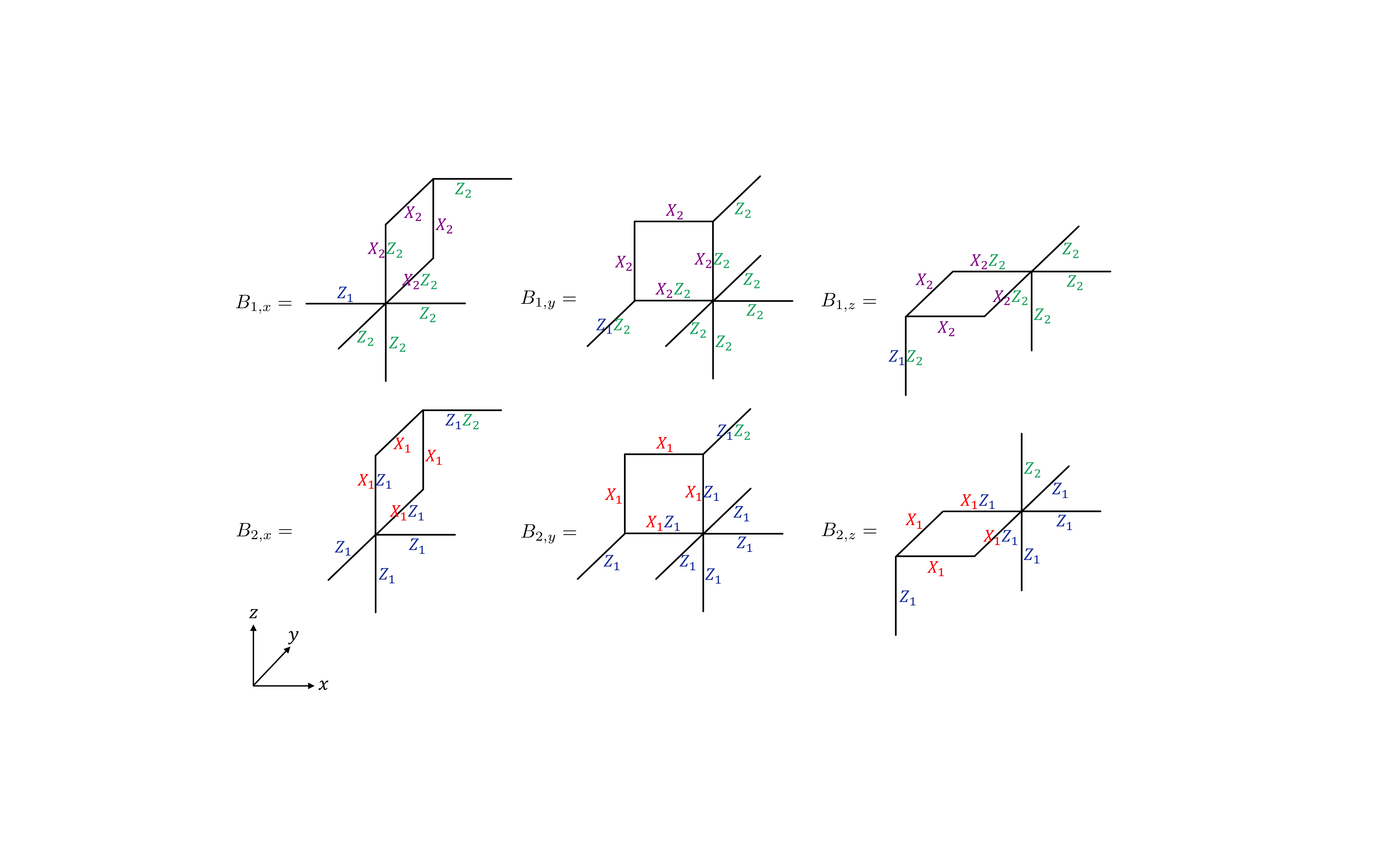}
    \caption{\label{fig: 3fWW_QCA_separator} The locally flippable separators used to define the QCA $\tilde{\alpha}_{3F}$. There are two qubits per edge with the Pauli $X$ and $Z$ operators of the first qubit denoted as ${\color{red} X_1}$ and ${\color{Blue} Z_1}$ and those of the second qubit represented by ${\color{violet} X_2}$ and ${\color{ForestGreen} Z_2}$.}
\end{figure*}

In this appendix, we construct a QCA $\tilde\alpha_{3F}$ belonging to the same class as the QCA $\alpha_{3F}$ of Ref.~\cite{HFH18} with the property that $\tilde{\alpha}_{3F}$ squares to the identity. To define $\tilde{\alpha}_{3F}$, we specify a locally flippable separator. We take the separators to be those shown in Fig.~\ref{fig: 3fWW_QCA_separator}.

Notice that the separators define a Hamiltonian that is a non-redundant version of the original 3-fermion Walker-Wang model of Ref.~\cite{WW12}.\footnote{More specifically, the vertex terms of the original construction are redundant. In our construction, vertex terms can be generated by the product of the plaquette terms, and therefore, the stabilizers are non-redundant.} 

To express the separators concisely, we introduce the following fermionic hopping operators $\tilde{U}_1$ and $\tilde{U}_2$ on edges~$e$:
\begin{eqs} \label{eq: fermionic hopping1}
    \tilde{U}_1=
    \vcenter{\hbox{\includegraphics[scale=.5,trim={0cm 0cm 0cm 0cm},clip]{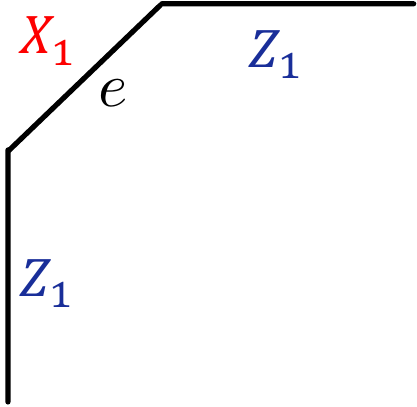}}},
    \vcenter{\hbox{\includegraphics[scale=.5,trim={0cm 0cm 0cm 0cm},clip]{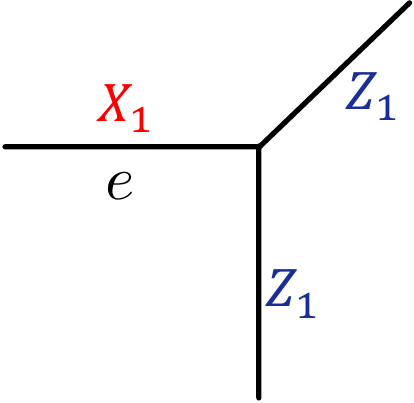}}}, 
    \vcenter{\hbox{\includegraphics[scale=.5,trim={0cm 0cm 0cm 0cm},clip]{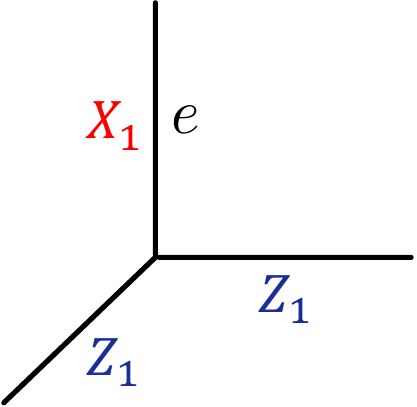}}},
\end{eqs}

\begin{figure*}[htb]
\centering
    \includegraphics[width=.75\textwidth, trim={0cm 0cm 0cm 0cm},clip]{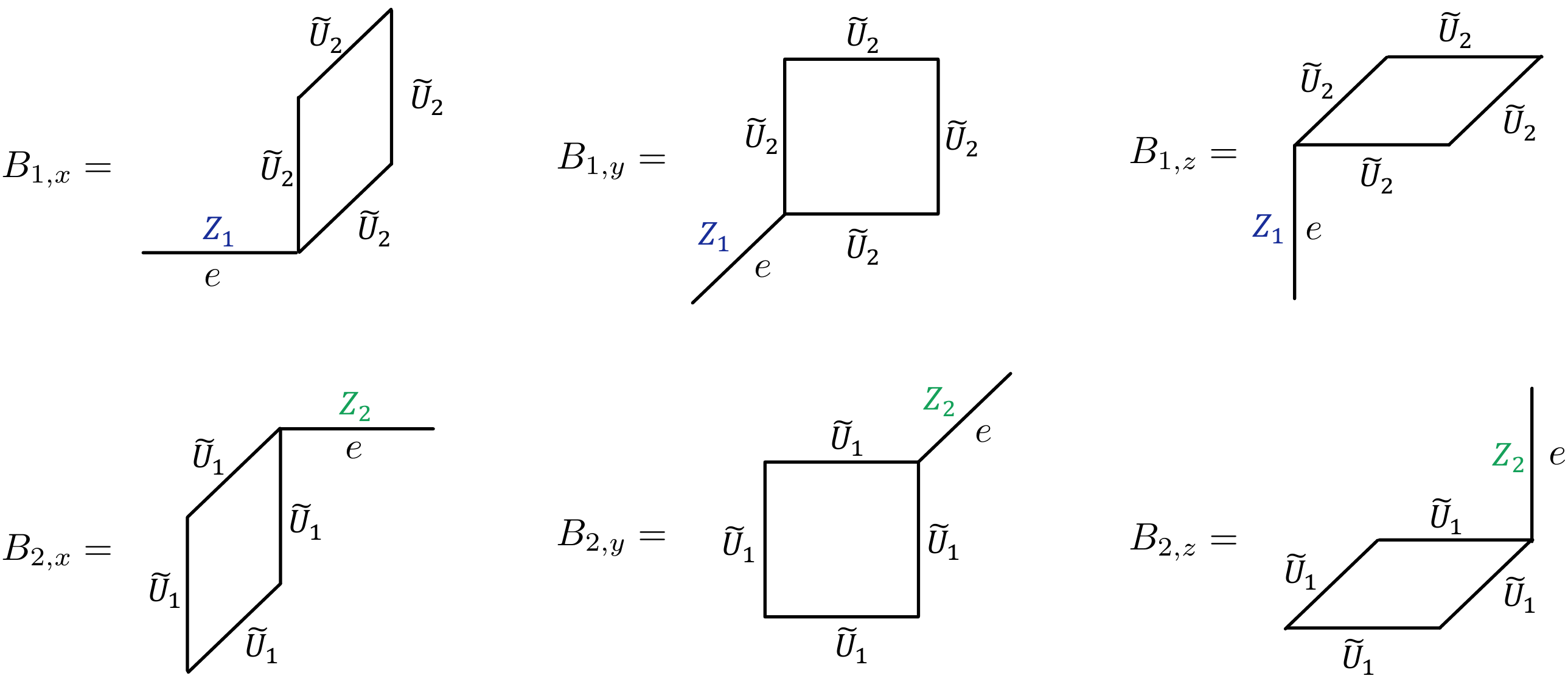}  \label{fig: separatorsU}
    \caption{The separators of Fig.~\ref{fig: 3fWW_QCA_separator} expressed using the fermionic hopping operators in Eqs.~\eqref{eq: fermionic hopping1} and \eqref{eq: fermionic hopping 2}. We notice that each separator is a single site Pauli operator multiplied by a loop of fermion hopping operators.}
\end{figure*}

\begin{eqs} \label{eq: fermionic hopping 2}
    \tilde{U}_2=
    \vcenter{\hbox{\includegraphics[scale=.5,trim={0cm 0cm 0cm 0cm},clip]{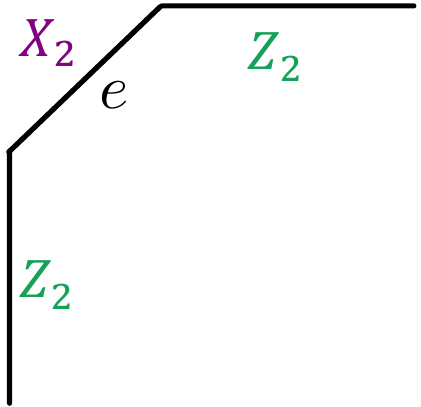}}},
    \vcenter{\hbox{\includegraphics[scale=.5,trim={0cm 0cm 0cm 0cm},clip]{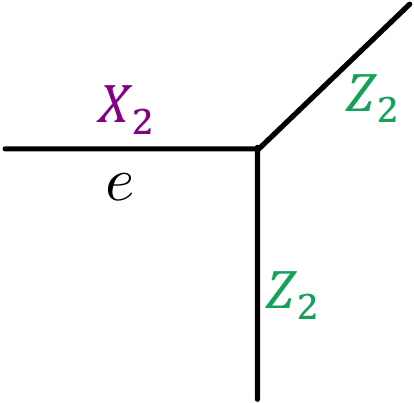}}}, 
    \vcenter{\hbox{\includegraphics[scale=.5,trim={0cm 0cm 0cm 0cm},clip]{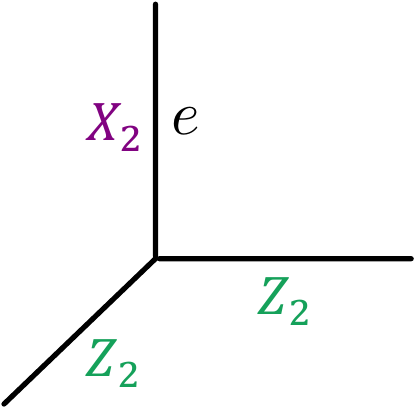}}}.
\end{eqs}
The separators in Fig.~\ref{fig: 3fWW_QCA_separator} can then be written more concisely as shown in Fig.~\ref{fig: separatorsU} (page 24).

Note that the representations of the separators in Fig.~\ref{fig: separatorsU} have sign ambiguities since $\tilde{U}_1$ and $\tilde{U}_2$ on different edges may anti-commute. We choose the convention that $Z$ operators are applied first, consistent with Fig.~\ref{fig: 3fWW_QCA_separator}. 

\begin{figure*}[htb]
    \centering
\includegraphics[width=.75\textwidth, trim={0cm 0cm 0cm 0cm},clip]{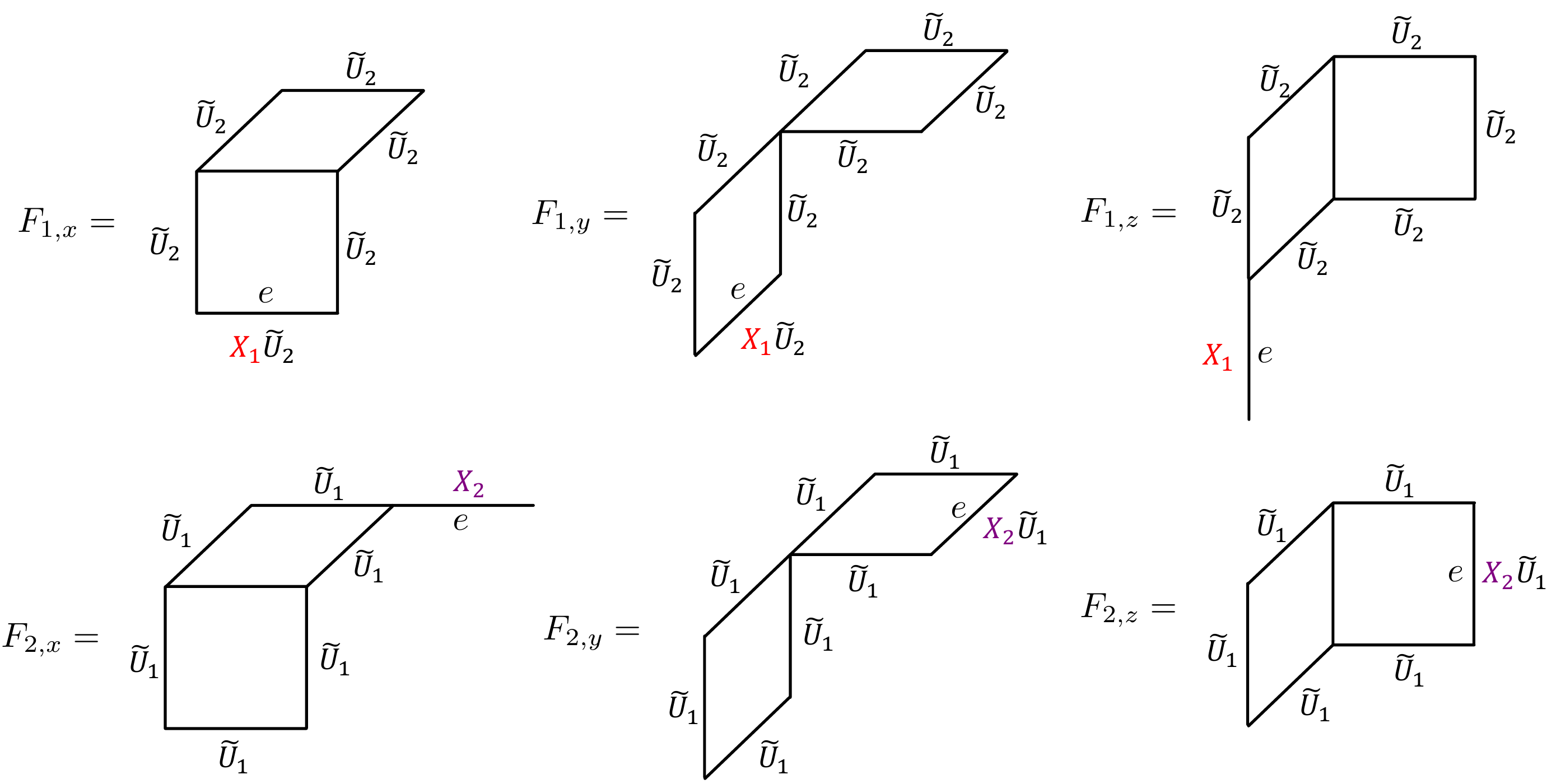} \label{fig: 3Fflippers}
    \caption{The mutually commuting flippers associated to the separators in Fig.~\ref{fig: separatorsU}. Again, we note that each flipper is equivalent to a single site Pauli operator multiplied by a closed string of hopping operators.}
\end{figure*}

We now define flippers as depicted in Fig~\ref{fig: 3Fflippers}. 
It can be checked that these flippers satisfy $2.$ of Definition~\ref{def:LFS} and have the additional property that they are mutually commuting.

Therefore, we can define the QCA $\tilde\alpha_{3F}$ by the mapping:
\begin{eqs} \label{eq: tildealpha3f def}
    \tilde\alpha_{3F}(B_1) &= Z_1, \qquad 
    \tilde\alpha_{3F}(B_2) = Z_2, \\
    \tilde\alpha_{3F}(F_1) &= X_1, \qquad
    \tilde\alpha_{3F}(F_2) = X_2, 
\end{eqs}
where $Z_1$ along the $x$-, $y$-, $z$-axes are mapped to $B_{1,x}$, $B_{1,y}$, $B_{1,z}$ respectively, and likewise for $X_1$, $Z_2$, and $X_2$.

To check the identity $\tilde\alpha_{3F}
^2 ( P) = P$ for any Pauli operator $P$, we first notice that the product of $\tilde{U}_1$ or $\tilde{U}_2$ around a face is invariant under $\tilde\alpha_{3F}$:
\begin{eqs}
    &\tilde\alpha_{3F} \left(\prod_{e \subset f} \tilde{U}_1 (e)\right) = \prod_{e \subset f} \tilde{U}_1 (e),\\
    &\tilde\alpha_{3F} \left(\prod_{e \subset f} \tilde{U}_2 (e)\right) = \prod_{e \subset f} \tilde{U}_2 (e).
\label{eq: product of U is invariant}
\end{eqs}
For example, we demonstrate this on the product of $\tilde{U}_1$ on a face in the $x$-direction. The product of $\tilde{U}_1$ around the face $f$ is
\begin{eqs}
    \prod_{e \subset f} \tilde{U}_1 (e) =
    \vcenter{\hbox{\includegraphics[scale=.4]{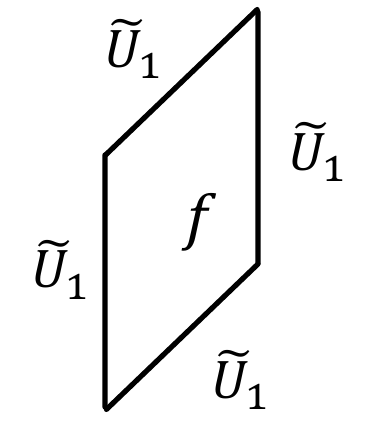}}}
    =
    \vcenter{\hbox{\includegraphics[scale=.4]{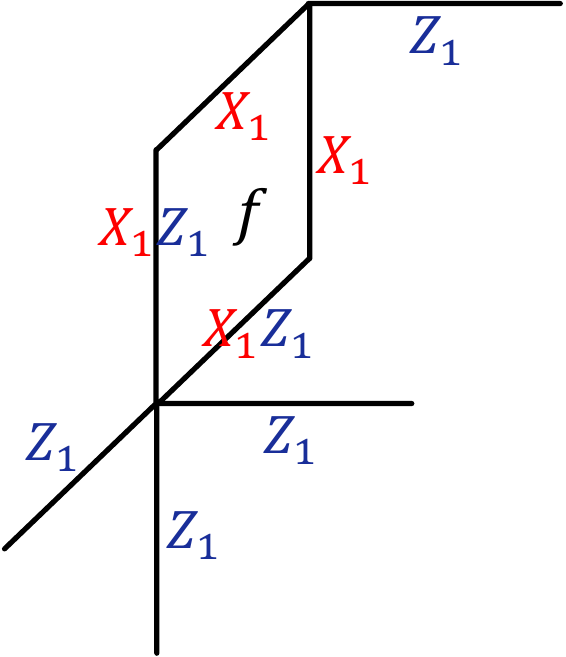}}}.
\end{eqs}

To clearly see how $\tilde\alpha_{3F}$ acts on the operator above, we treat the $X$ part and the $Z$ part separately. The $X$ operators are mapped as:
\begin{eqs}
    \vcenter{\hbox{\includegraphics[scale=.42]{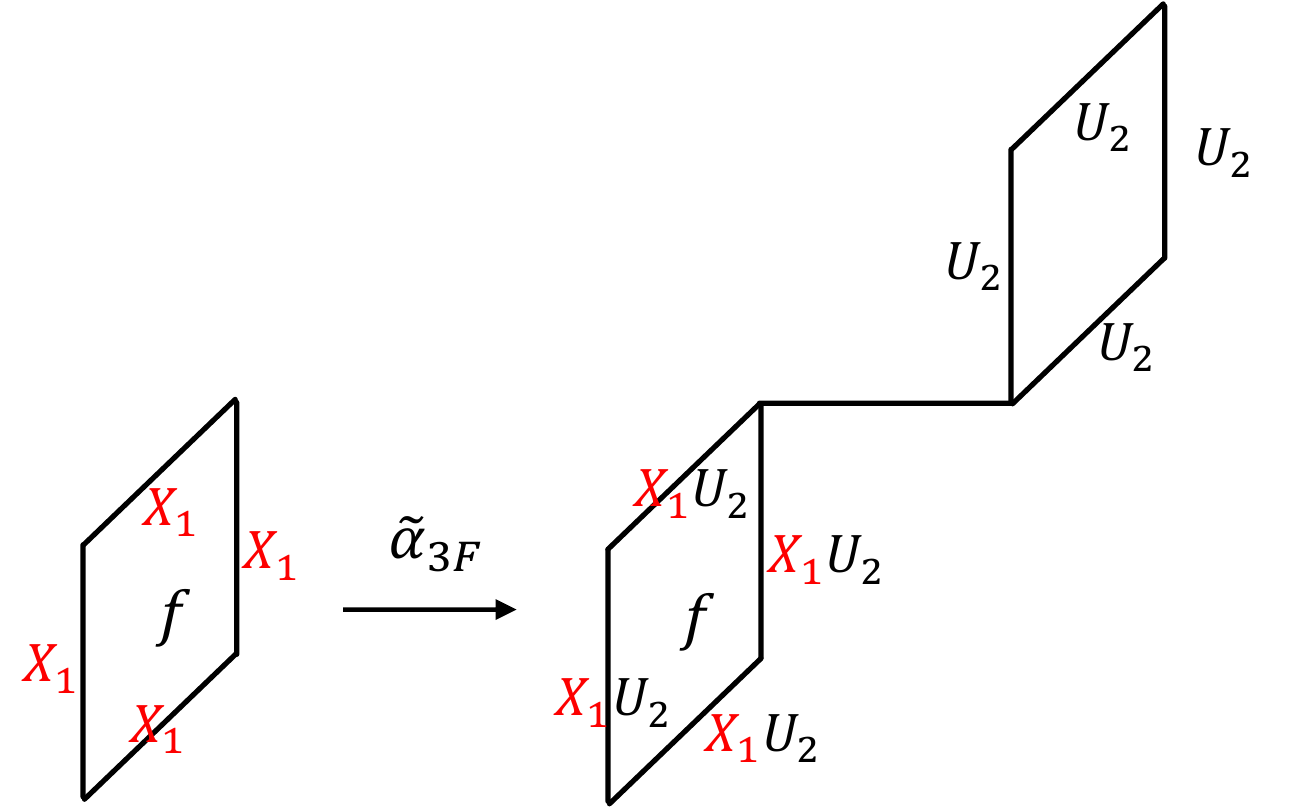}}},
\end{eqs}
while the $Z$ operators are mapped as:
\begin{eqs}
    \vcenter{\hbox{\includegraphics[scale=.45]{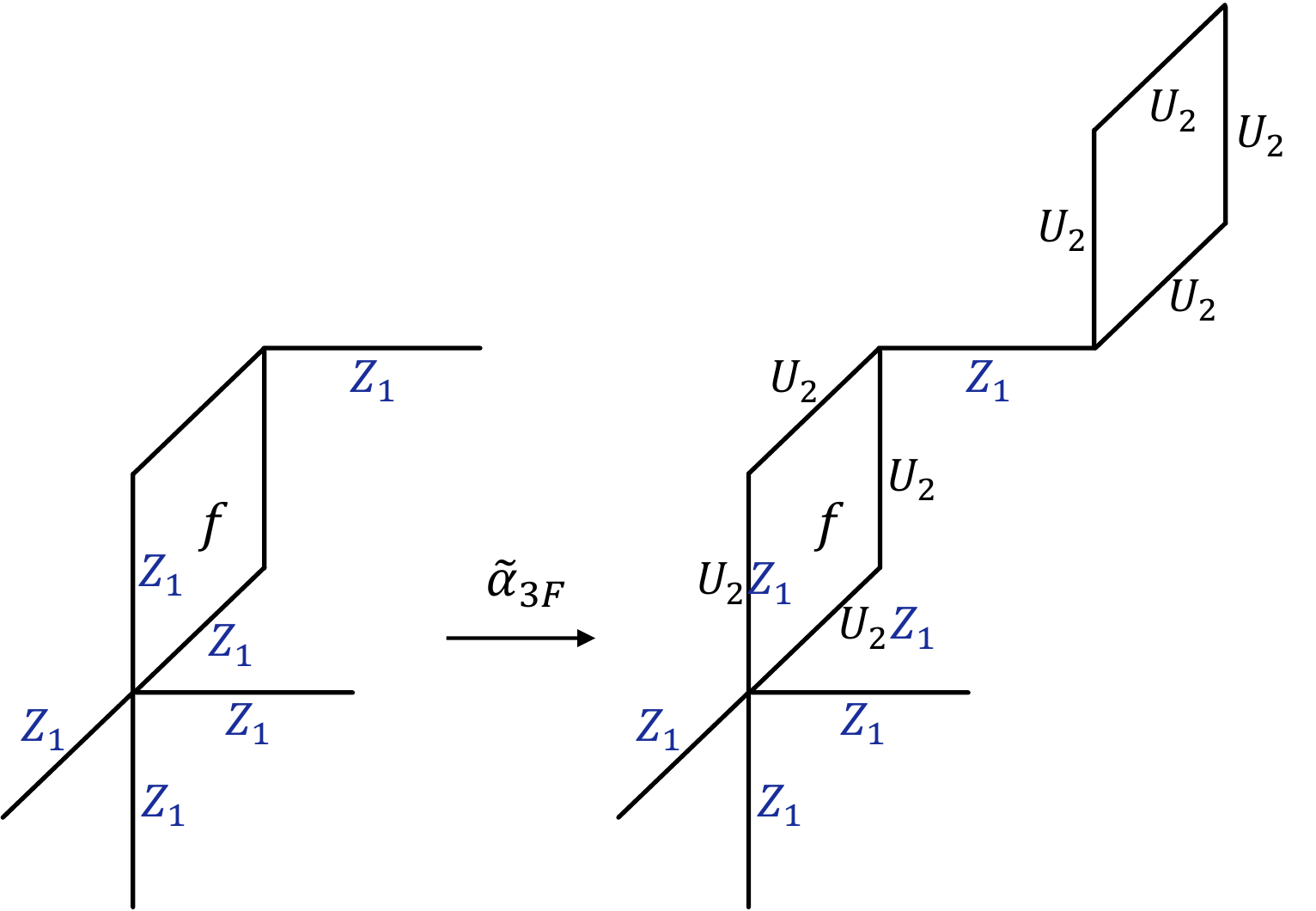}}}.
\end{eqs}
The products of $\tilde{U}_2$ in the $X$ part and the $Z$ part cancel each other. Since our convention is to apply the $Z$ part first and the $X$ part second, the $\tilde{U}_2$ loops cancel directly without giving any extra sign. Therefore, we have proved Eq.~\eqref{eq: product of U is invariant} for one case. All other cases can be checked in a similar manner.

Given that the product of $\tilde{U}_1$ or $\tilde{U}_2$ on a closed loop is invariant under $\tilde\alpha_{3F}$, we can immediately conclude that $\tilde\alpha_{3F}^2 = 1$ since the QCA defined by Eq.~\eqref{eq: tildealpha3f def} simply attaches loops of $\tilde{U}_1$ and $\tilde{U}_2$ to a single Pauli operator. $\tilde{U}_1$ or $\tilde{U}_2$ square to the identity, so if we apply $\tilde\alpha_{3F}$ twice, it must be the identity. This can also be confirmed by using polynomials. The QCA $\tilde{\alpha}_{3F}$ corresponds to the symplectic matrix below, which squares to the $12 \times 12$ identity matrix. Explicitly, we let $\tilde\alpha_{3F}$ be the matrix $\begin{pmatrix} B_1 & B_2 & F_1 & F_2 \end{pmatrix}$,
where $B_{1}$, $B_2$, $F_{1}$, and $F_2$ are the $12 \times 3$ matrices (see supplementary Mathematica file):
\onecolumngrid
\begin{align*}
    B_1=
    \begin{pmatrix}
        1 & 0 & 0  \\
        0 & 1 & 0  \\
        0 & 0 & 1  \\
        x y z+x & x y+y & x y z+y z  \\
        \frac{x}{y}+x & x y z+x y+x+1 & x y z+x z \\
        \frac{x}{z}+x & \frac{x y}{z}+x y & x y+1   \\
        0 & 0 & 0   \\
        0 & 0 & 0   \\
        0 & 0 & 0  \\
        0 & y z+y & y z+z   \\
        x z+x & 0 & x z+z   \\
        x y+x & x y+y & 0   
    \end{pmatrix}, 
    \quad
    B_2=
    \begin{pmatrix}
        \frac{1}{y z}+1 & \frac{1}{x z}+\frac{1}{z} & \frac{1}{x}+1 \\
        \frac{1}{y^2 z}+\frac{1}{y z} & \frac{1}{x y z}+\frac{1}{y z}+\frac{1}{z}+1 & \frac{1}{y}+1 \\
        \frac{1}{y z^2}+\frac{1}{y z} & \frac{1}{z^2}+\frac{1}{z} & \frac{1}{x y z}+\frac{1}{z} \\
        1 & 0 & 0 \\
        0 & 1 & 0\\
        0 & 0 & 1  \\
        0 & \frac{1}{x z}+\frac{1}{x} & \frac{1}{x y}+\frac{1}{x}  \\
        \frac{1}{y z}+\frac{1}{y} & 0 & \frac{1}{x y}+\frac{1}{y}  \\
        \frac{1}{y z}+\frac{1}{z} & \frac{1}{x z}+\frac{1}{z} & 0 \\
        0 & 0 & 0   \\
        0 & 0 & 0   \\
        0 & 0 & 0  
    \end{pmatrix},
\end{align*}
\begin{align*}
    F_1=
    \begin{pmatrix}
        0 & 0 & 0  \\
        0 & 0 & 0  \\
        0 & 0 & 0  \\
        x y z+x+y z+1 & x y^2 z+y^2 z+y z+1 & x y z+y z^2+y z+z  \\
        x y z+\frac{x}{y}+x+\frac{1}{y} & x y^2 z+x y z+\frac{1}{y}+1 & x y z^2+x y z+x z+\frac{z}{y} & \\
        x y+\frac{x}{z}+x+1 & x y^2+y+\frac{1}{z}+1 & x y z+x y+z+1 \\
        1 & 0 & 0  \\
        0 & 1 & 0  \\
        0 & 0 & 1  \\
        y z+1 & y^2 z+y z & y z^2+y z  \\
        x z+z & x y z+y z+z+1 & z^2+z  \\
        x+1 & y+1 & x y z+z 
    \end{pmatrix},
\end{align*}    
\begin{align*}
    F_2=
    \begin{pmatrix}
        \frac{1}{x y z}+\frac{1}{x}+\frac{1}{y z}+1 & \frac{1}{x y z}+\frac{y}{x}+\frac{1}{x}+y & \frac{1}{x y}+\frac{z}{x}+\frac{1}{x}+1 \\
        \frac{1}{x y^2 z}+\frac{1}{y^2 z}+\frac{1}{y z}+1 & \frac{1}{x y^2 z}+\frac{1}{x y z}+y+1 & \frac{1}{x y^2}+\frac{1}{y}+z+1 \\
        \frac{1}{x y z}+\frac{1}{y z^2}+\frac{1}{y z}+\frac{1}{z} & \frac{1}{x y z^2}+\frac{1}{x y z}+\frac{1}{x z}+\frac{y}{z} & \frac{1}{x y z}+\frac{1}{x y}+\frac{1}{z}+1 \\
        0 & 0 & 0 \\
        0 & 0 & 0 \\
        0 & 0 & 0 \\
        \frac{1}{x y z}+\frac{1}{x} & \frac{y}{x}+\frac{1}{x} & \frac{z}{x}+\frac{1}{x} \\
        \frac{1}{x y}+\frac{1}{y} & \frac{1}{x y z}+\frac{1}{x y}+\frac{1}{x}+1 & \frac{z}{x y}+\frac{1}{x y} \\
        \frac{1}{x y z}+\frac{1}{y z} & \frac{1}{x y z}+\frac{1}{x z} & \frac{1}{x y}+1 \\
        1 & 0 & 0 \\
        0 & 1 & 0 \\
        0 & 0 & 1
    \end{pmatrix}.
\end{align*}

\end{document}